\documentclass[fleqn,usenatbib]{mnras}

\usepackage{newtxtext,newtxmath}

\usepackage[T1]{fontenc}

\DeclareRobustCommand{\VAN}[3]{#2}
\let\VANthebibliography\thebibliography
\def\thebibliography{\DeclareRobustCommand{\VAN}[3]{##3}\VANthebibliography}

\usepackage{graphicx}	
\usepackage{amsmath}	

\usepackage{amssymb}	

\defcitealias{2012MNRAS.426..140D}{DVS12}

\title[The importance of how SN energy is distributed around stars]{The importance of the way in which supernova energy is distributed around young stellar populations in simulations of galaxies}

\author[E. Chaikin et al.]{Evgenii Chaikin,$^{1}$\thanks{E-mail: chaikin@strw.leidenuniv.nl},
Joop Schaye,$^{1}$
Matthieu Schaller,$^{2,1}$
Yannick M. Bah\'{e},$^{1}$ \newauthor
Folkert S. J. Nobels$^{1}$ and
Sylvia Ploeckinger$^{2}$
\\
$^{1}$Leiden Observatory, Leiden University, PO Box 9513, 2300 RA Leiden, The Netherlands\\
$^{2}$Lorentz Institute for Theoretical Physics, Leiden University, PO Box 9506, 2300 RA Leiden, The Netherlands
}

\date{Accepted XXX. Received YYY; in original form ZZZ}

\pubyear{2015}

\begin{document}
\label{firstpage}
\pagerange{\pageref{firstpage}--\pageref{lastpage}}
\maketitle

\begin{abstract}
Supernova (SN) feedback plays a crucial role in simulations of galaxy formation. Because blastwaves from individual SNe occur on scales that remain unresolved in modern cosmological simulations, SN feedback must be implemented as a subgrid model. Differences in the manner in which SN energy is coupled to the local interstellar medium and in which excessive radiative losses are prevented have resulted in a zoo of models used by different groups. However, the importance of the selection of resolution elements around young stellar particles for SN feedback has largely been overlooked. In this work, we examine various selection methods using the smoothed particle hydrodynamics code \textsc{swift}. We run a suite of isolated disk galaxy simulations of a Milky Way-mass galaxy and small cosmological volumes, all with the thermal stochastic SN feedback model used in the \textsc{eagle} simulations. We complement the original mass-weighted neighbour selection with a novel algorithm guaranteeing that the SN energy distribution is as close to isotropic as possible. Additionally, we consider algorithms where the energy is injected into the closest, least dense, or most dense neighbour. We show that different neighbour-selection strategies cause significant variations in star formation rates, gas densities, wind mass loading factors, and galaxy morphology. The isotropic method results in more efficient feedback than the conventional mass-weighted selection. We conclude that the manner in which the feedback energy is distributed among the resolution elements surrounding a feedback event is as important as changing the amount of energy by factors of a few.
\end{abstract}




\section{Introduction}
\label{sec: introduction}

Supernova (SN) feedback plays a vital role in galaxy formation and evolution. Without SN feedback, star formation occurs on a free-fall time-scale in a run-away fashion and galaxies are produced with too large stellar masses, both of which contradict observations \citep[e.g][]{WF1991,2012ApJ...745...69K,2017ApJ...846...71L}. Conversely, the inclusion of SN feedback enables galaxies to self-regulate, form stars at an expected rate, and maintain realistic morphologies throughout cosmic time \citep[e.g][]{2015MNRAS.446..521S, Dubois2015, Pillepich2018,Hopkins2018Fire2}. SN feedback is crucial for enriching the interstellar medium (ISM) with metals \citep[e.g.][]{2001ApJ...561..521A, 2011MNRAS.415..353W} as well as maintaining the multiphase structure of the ISM \citep[e.g.][]{McKeeOstriker1977}. 

SNe inject energy and momentum into the ISM on a scale of $\sim 10-10^2$ pc \citep[e.g.][]{2015ApJ...802...99K}. Although such scales can be probed in idealised simulations of dwarf galaxies \cite[e.g.][]{2017MNRAS.471.2151H,2017MNRAS.470L..39F,Gutcke2021}, in large cosmological simulations they remain significantly below the current resolution limit, which in state-of-the-art cosmological simulations of representative volumes is typically $\sim 10^2 - 10^3$ pc \citep[e.g.][]{2015MNRAS.446..521S,Pillepich2018}. In the latter case, the evolution of SN remnants cannot be accurately followed either in time or in space - instead, a subgrid model for SN feedback is adopted. Such subgrid models normally have one or several free parameters, which require calibration in order to produce a realistic galaxy population \citep[e.g.][]{Crain2015}. Although at $\sim10$ pc and higher resolution all `good' SN subgrid models are expected to converge to the same answer\footnote{This is because the Sedov-Taylor phase of the SN blast evolution becomes well-resolved at most gas densities in the simulation(s).} \citep[e.g.][]{Smith2018}, the degree of uncertainty as to which model is most suitable for simulations at much lower resolution remains very high \citep[e.g.][]{RosdahlSchaye2017} and generally depends on the problem being studied.

In the classic and arguably simplest subgrid model for SN feedback implemented in a smoothed particle hydrodynamics (SPH) code, the canonical $10^{51}$ erg of energy per SN event is directly injected into the gas neighbours of the stellar particle in thermal form \citep{Katz1992}. At the resolution of cosmological simulations, this approach always leads to SN feedback that is too inefficient because the injected energy is smoothed over too much mass and is inevitably radiated away too quickly. The aforementioned failure of SN feedback has been known for several decades under the name `overcooling problem' \citep[e.g.][]{1996ApJS..105...19K}. 

Alternatively, the energy from SNe can be injected into the gas elements in kinetic form \citep[e.g][]{NavarroWhite1993,2001ApJ...558..598K} where the desired amount of momentum from the SN blast is added to the gas by applying `a kick' to the gas neighbours (usually in a direction away from the stellar particle). By kicking gas particles with sufficiently high velocities, one can reduce the energy losses due to overcooling: for higher kick velocities, a larger fraction of the injected (kinetic) energy will fuel galactic winds and less energy will be thermalised and subsequently lost to radiation. The kinetic model is often complemented by temporarily switching off hydrodynamic forces for the kicked particles \citep{SpringelHernquist2003,OppenheimerDave2006}, which can dramatically strengthen the feedback \citep{DallaVecchiaSchaye2008}, can improve convergence with resolution, and allows one to attain desired galaxy-scale wind mass loading factors by construction. The wind particles are recoupled to the hydrodynamics after they have escaped the dense, star-forming phase, which is estimated by the moment when their local density falls below a certain value or a maximum travel time has been reached \citep{SpringelHernquist2003, 2013MNRAS.436.3031V}. Drawbacks of decoupled winds are that they may implicitly use much more energy than is specified and that they cannot generate turbulence or blow bubbles in the ISM. 

The second common approach to make SN feedback efficient is to keep injecting energy in thermal form but manually disable radiative cooling of the heated gas particles for a certain period of time \citep{Stinson2006,Teyssier2013,Dubois2015}. The use of such `\textit{delayed cooling}' models is sometimes justified by noting that some of the SN energy should in reality be attributed to (non-thermal) processes with (much) longer cooling time-scales that cannot be resolved in numerical simulations due to the lack of physics and resolution. Drawbacks of delayed cooling are that it tends to result in excessive amounts of gas with short cooling times, and hence may predict excessive emission and absorption by commonly observed species, and that there is no clear separation in scale between subgrid and resolved processes. Similar by nature are \textit{multi-phased} medium SN models \citep[e.g.][]{Scannapieco2006,Keller2014}. Briefly, in such models, the hot and cold ISM phases are traced separately; storing (part of) the SN energy in the hot, tenuous phase within a resolution element reduces the cooling losses. Drawbacks of this approach are that one requires a semi-analytic model of the multi-phase gas and that this model will even be active at resolved densities. 

Another important class of SN subgrid models is that based on the evolution of the SN blast itself. It is well known that the blast momentum increases by more than an order of magnitude during the energy conserving phase \cite[e.g.][]{2015ApJ...802...99K}. If this phase cannot be resolved, a boost to the momentum can be applied by hand to obtain the value expected from theory. Such models are called \textit{mechanical} \citep{Kimm2014,hopkinsfeedback2018,Marinacci2019}. A drawback is that such models do not correct for excessive cooling losses downstream, e.g. when different bubbles collide, and therefore tend to require higher resolution (or more energy) than alternative approaches.

Lastly, one can inject thermal SN energy stochastically (\citealt{2012MNRAS.426..140D}, from henceforth \citetalias{2012MNRAS.426..140D}).  In this model, the amount of energy released per single SN feedback event is a free parameter, which allows one to reduce the radiation losses as much as needed by increasing the energy per feedback event. Such models are referred to as \textit{stochastic} in the sense that for large energies per feedback event, the decision on whether a gas neighbour receives this energy or not is probabilistic. In particular, in the \citetalias{2012MNRAS.426..140D} model, a stellar particle heats its gas neighbours to a predefined temperature $\Delta T$ with the probability inversely proportional to $\Delta T$. The heating temperature $\Delta T$ is usually quite high, to overcome numerical overcooling: in the \textsc{eagle} simulation \citep{2015MNRAS.446..521S}, for example, it was set to $10^{7.5}$ K. The drawback is then that the wind may be overly hot and the bubbles too large, at least close to the resolution limit \citep[e.g.][]{2016MNRAS.456.1115B}.

The evolution of galaxies in numerical simulations is affected not only by the internal design choices of SN subgrid models discussed above, but also by the assumed delay between star formation and feedback \citep[e.g.][]{2020arXiv200403608K} and the degree of clustering of SNe \citep[e.g.][]{2014MNRAS.443.3463S,Gentry2017}, all of which are generally resolution dependent. Moreover, there is uncertainty in how gas elements that receive the feedback energy are selected, which can affect the densities at which the explosion energy is deposited; the environment where SNe go off largely determines the structure of the ISM \citep[e.g.][]{2015MNRAS.449.1057G,2016MNRAS.456.3432G}.

Generally, it is desirable for any SN feedback model to be statistically isotropic \citep[e.g.][]{hopkinsfeedback2018,2019MNRAS.483.3363H}. Failing to satisfy this requirement might generate unphysical shells sweeping up most of the ISM gas in disk galaxies and destroying their otherwise stable disks \citep{Smith2018}. In simulations with SPH codes, SN energy is often distributed proportionally to the kernel function of the stellar particle that does feedback, evaluated at the separation between the star and its gas neighbours \citep[e.g.][]{Scannapieco2006,Stinson2006,2013MNRAS.428..129S}. Such weighting schemes can be viewed as \textit{mass-weighted} rather than isotropic; that is because solid angles with more gas particles (and hence more mass) on average collect more energy. Although extensive research has been done comparing different prescriptions for SN subgrid models  \citep[e.g.][]{2010MNRAS.402.1536S,2012MNRAS.423.1726S, RosdahlSchaye2017, 2017MNRAS.470.3167V, Smith2018, Gentry2020,AGORA2021}, no systematic work exists on addressing the impact of choices of gas-element selection for SN feedback. 

\begin{figure*}
 \centering
  \includegraphics[width=0.31\textwidth]{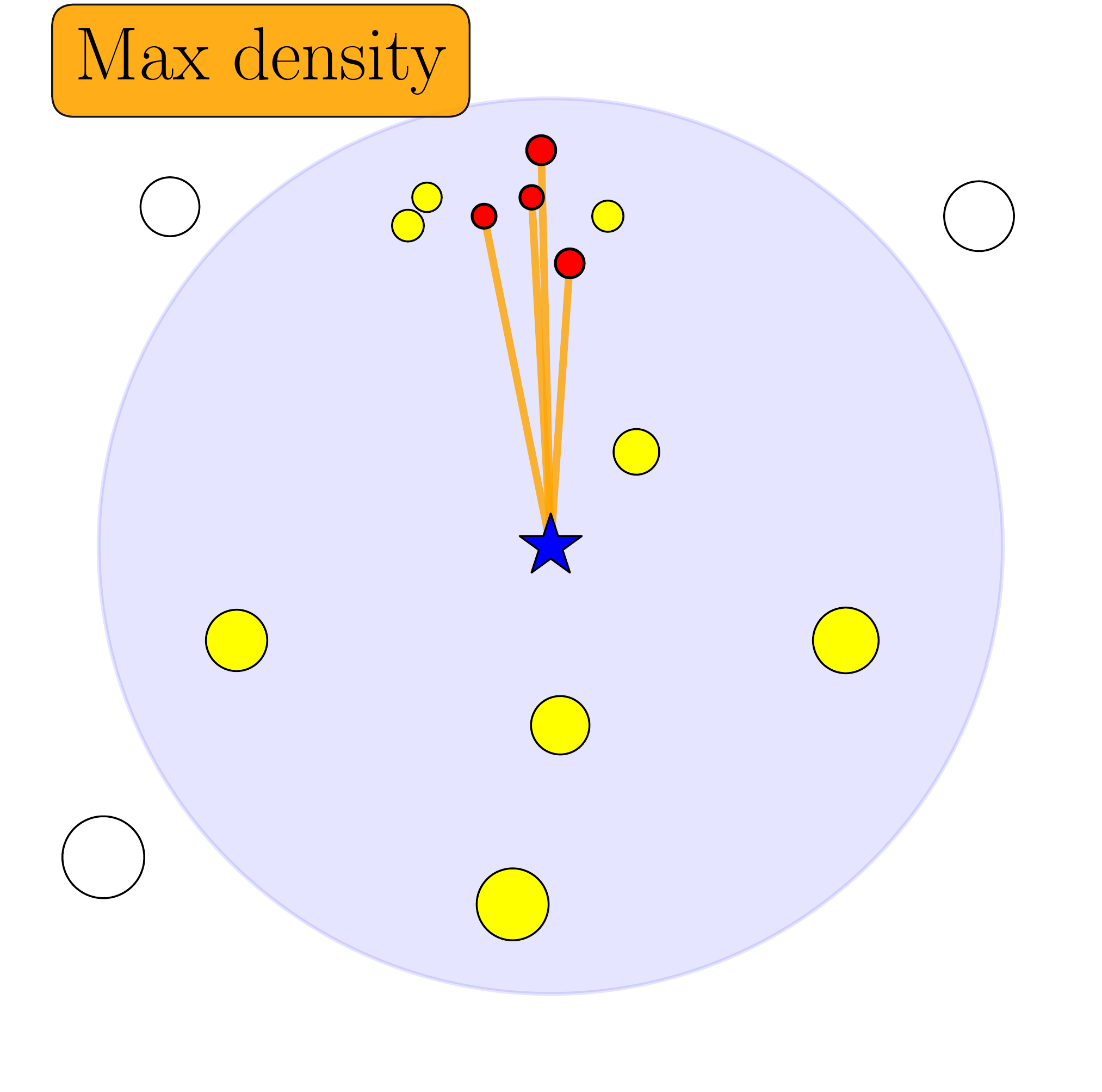}
  \includegraphics[width=0.31\textwidth]{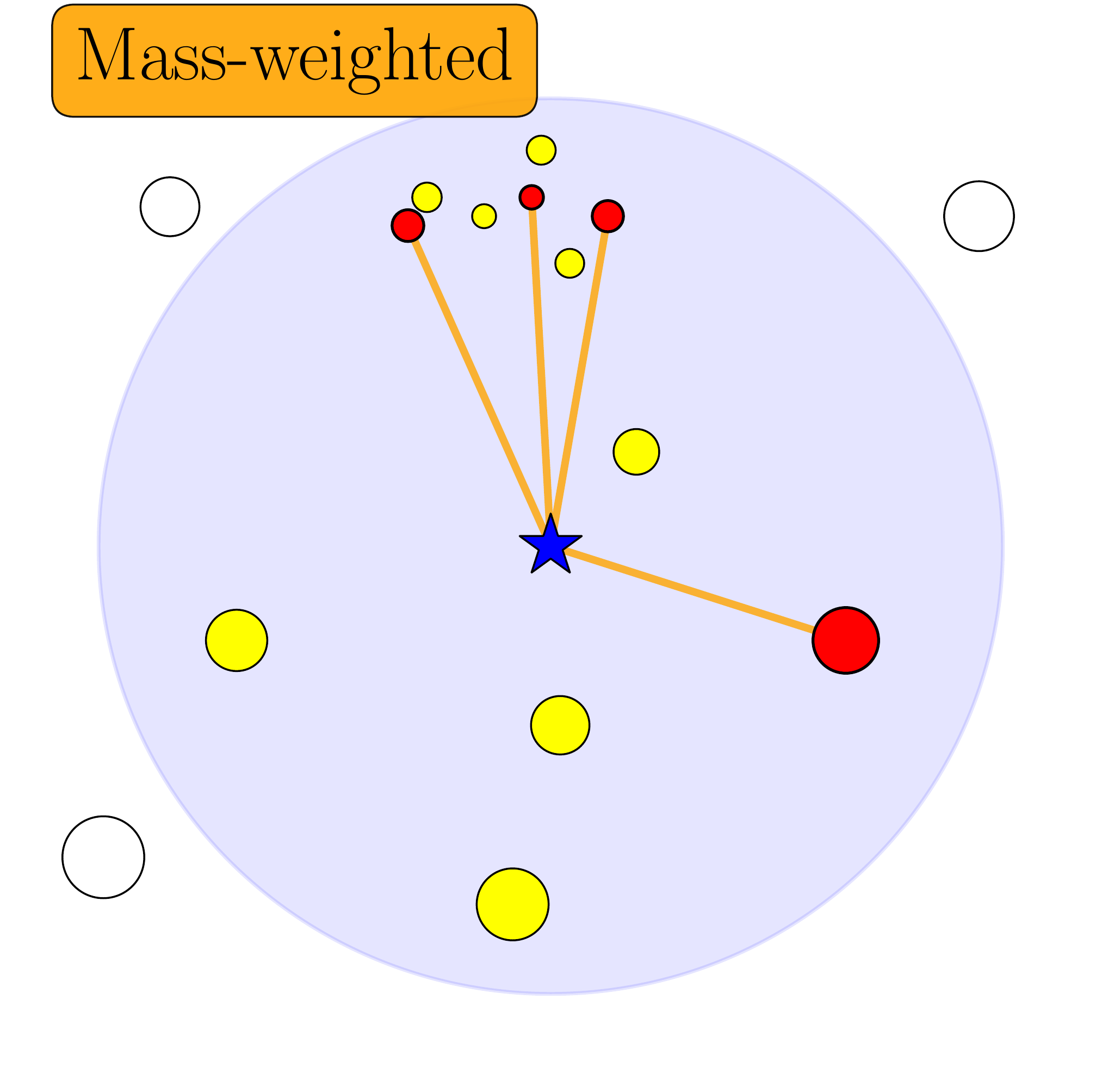}
  \includegraphics[width=0.31\textwidth]{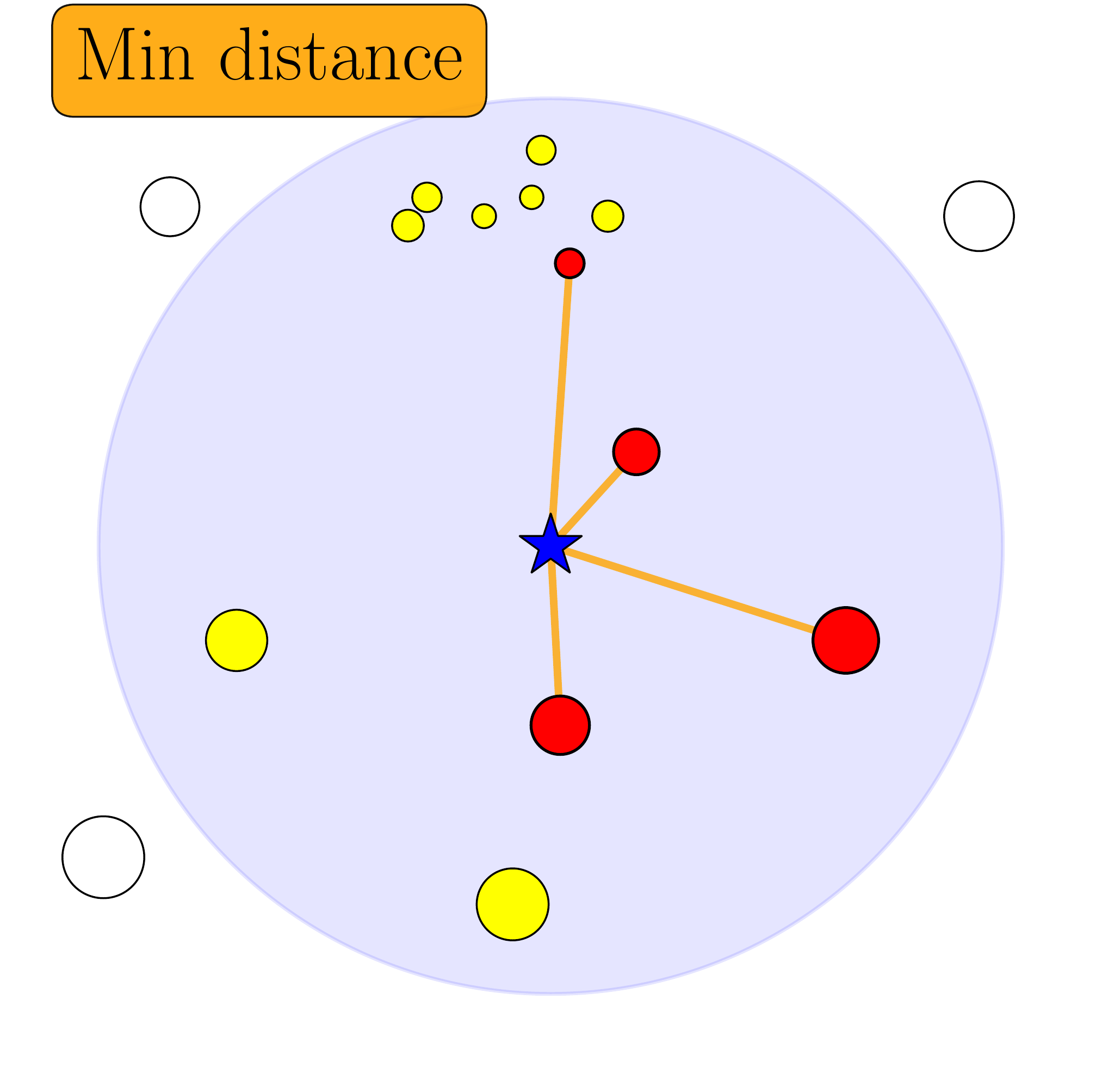}
  \includegraphics[width=0.31\textwidth]{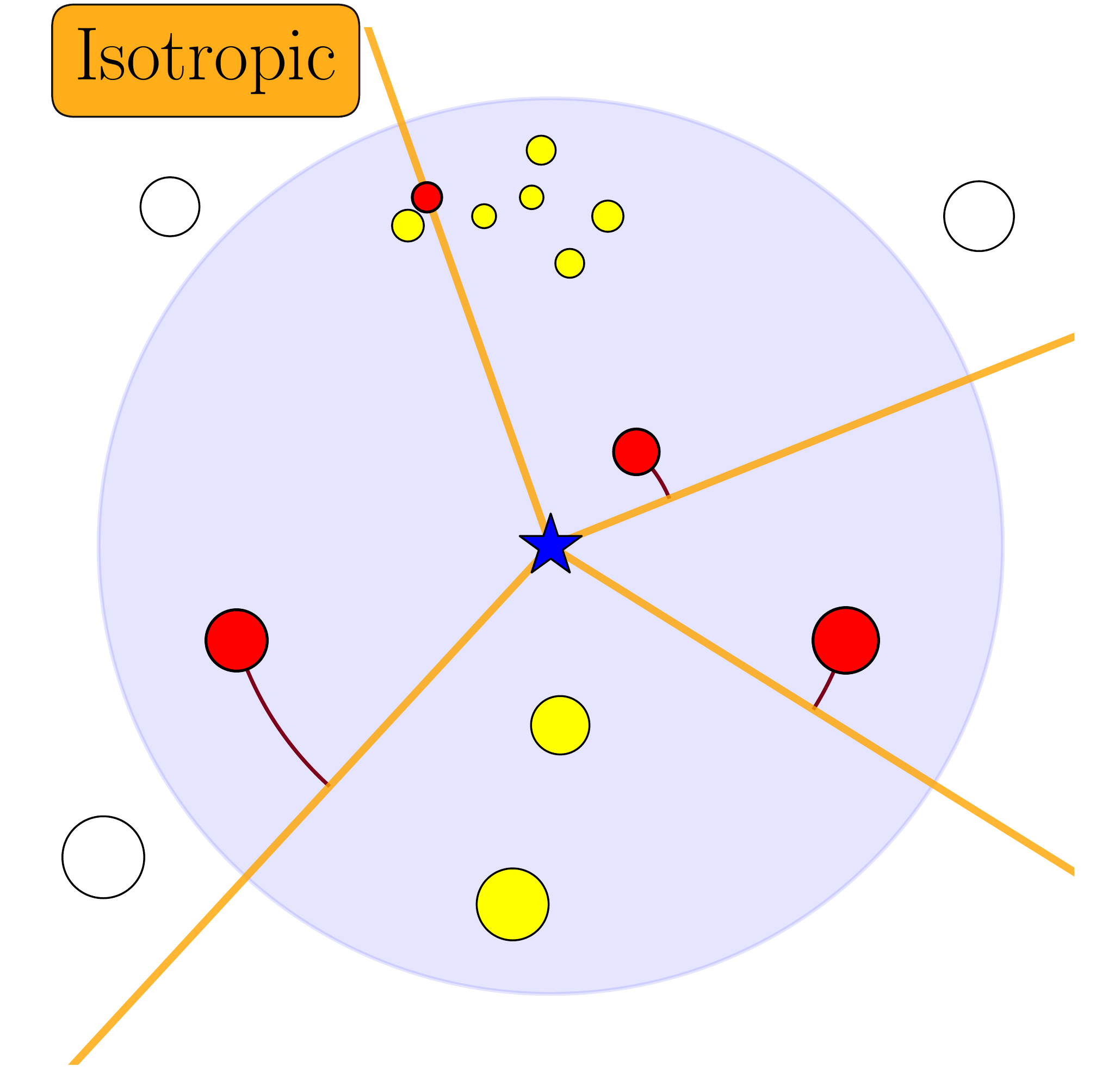}
  \includegraphics[width=0.31\textwidth]{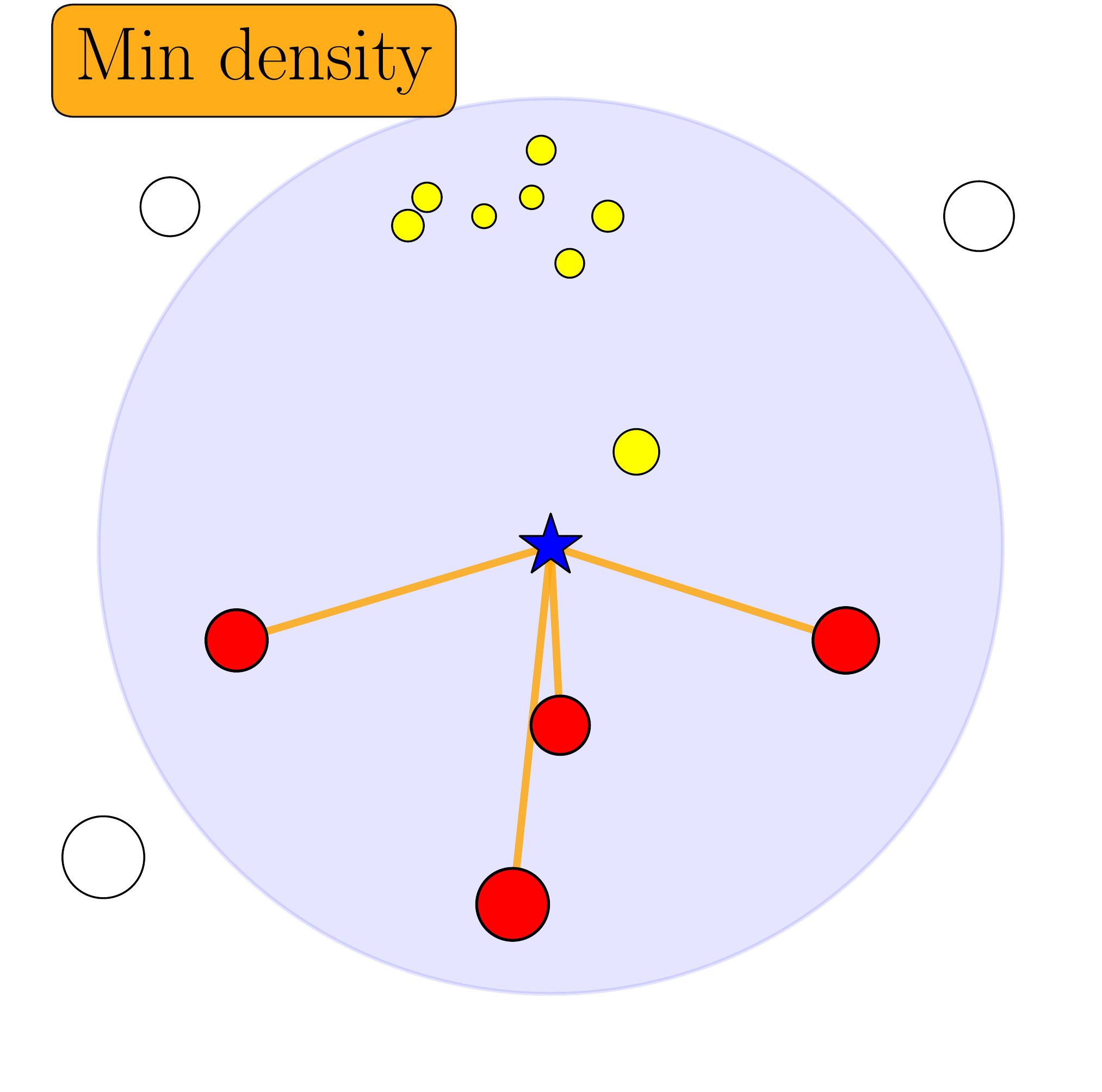}
  
    \caption{Five methods of gas-neighbour selection for energy injection in SN feedback, implemented in an SPH scheme. The shaded circular regions represent an SPH kernel of the stellar particle depicted by the blue star in the centre. Its gas neighbours are shown as yellow (and red) circles, whereas white circles represent gas particles outside the kernel. The size of the circles increases with decreasing density of the corresponding gas particles. We assume that the stellar particle does SN feedback by injecting \textit{four times} a fixed amount of SN energy into its gas neighbours. The gas neighbours that receive this energy are coloured red (otherwise they remain yellow). Depending on the neighbour-selection method, different gas neighbours will be selected to receive the energy. We consider five cases: \textsf{max\_density}, \textsf{mass-weighted},  \textsf{min\_distance}, \textsf{isotropic}, and \textsf{min\_density} (see text for details). The black arc lengths in the isotropic case (bottom left) are shown for clarity.}
    \label{fig:feedback_models}
\end{figure*}

In this work, we quantify the importance of the way in which gas elements are selected for the injection of SN energy, and study how strongly galaxy properties are affected by the variations in gas-element-selection models.  We run isolated disk galaxy simulations as well as simulations of small cosmological volumes. To the best of our knowledge, this is the first work focused on studying the effects of gas-neighbour selection. To minimize the number of free parameters, we keep the internal design of the SN feedback model fixed: in all our tests, we use the \citetalias{2012MNRAS.426..140D} thermal stochastic model with a fixed heating temperature and energy, and only vary the way in which gas elements are selected for SN feedback. In Section \ref{sec: neighbour_selection}, we describe five algorithms -- including the original one from \citetalias{2012MNRAS.426..140D} -- of how gas particles can be chosen for receiving SN energy. In Section \ref{sec: numerical simulations}, we describe the numerical simulations used in this work. Finally, in Section \ref{sec: Results}, we present the results of the simulations, which is followed by the discussion (Section \ref{sec: discussion}) and conclusions (Section \ref{sec: conclusions}).

\section{Neighbour selection}
\label{sec: neighbour_selection}

In this section, we consider a common situation when modelling SN feedback where, in a given time-step, a star particle with $N_{\rm ngb}$ gas elements in its kernel does SN feedback by injecting energy into the surrounding gas in $N_{\rm inj}$ equal chucks, each of energy $\Delta E_{\rm inj}$. We introduce five neighbour-selection prescriptions whose purpose is to determine how the $N_{\rm inj}$ energy injections will be distributed among the $N_{\rm ngb}$ gas neighbours. 

\subsection{Mass-weighted method}
\label{sec: neighbour_selection_mw}
 
In SPH, the most straightforward and most commonly used method for neighbour selection is to give all gas neighbours \textit{equal} probabilities to be selected for SN energy injection. If there is a density gradient inside the stellar kernel, this implies that more gas neighbours will be selected for SN feedback from the denser region than from the other parts of the kernel's volume. That is because the denser region is made up of a greater number of gas elements, which is a direct consequence of the fact that SPH is a Lagrangian method. As a result, most of the SN energy will be received by the denser gas, even though this denser gas may only comprise a small fraction of the kernel's total volume. 

From here on we will refer to this neighbour-selection method as the \textsf{mass-weighted} method.  We also note that this is the method that was used by \citetalias{2012MNRAS.426..140D}. 
 
\subsection{Isotropic method}
\label{subsectuion isotropic model}

To resolve the density bias outlined in \S\ref{sec: neighbour_selection_mw}, we now introduce a new neighbour-selection algorithm ensuring that the SN energy injections are statistically distributed isotropically as seen by the star\footnote{Here \textit{isotropic} means that the probability density distributions of the polar angle $\varphi$ and the cosine of the azimuthal angle $\theta$ in spherical coordinates, centred on the position of the stellar particle that does SN feedback, are uniform in the limit of an infinite number of gas neighbours in the kernel.}.

Given a stellar particle with $N_{\rm ngb}$ gas neighbours that injects SN energy $N_{\rm inj}$ times, where each injection event is of energy $\Delta E_{\rm inj}$, we take the following steps:

\begin{enumerate}

    \item  We cast $N_{\rm rays}$ rays in different, randomly chosen (i.e. uniform probability in $\varphi$ and $\cos \theta$ in spherical coordinates) directions from the position of the star;

    \item  For each of the $N_{\rm rays}$ rays, we calculate great-circle distances on a unit sphere between the ray and each of the $N_{\rm ngb}$ gas neighbours by using the haversine formula 
\begin{equation}
 \Omega_{ij} = 2\arcsin \sqrt{\sin^2\left(\frac{\theta_{j}-\theta_{i}}{2}\right) + \cos(\theta_{i})\cos(\theta_{j})\sin^2\left(\frac{\varphi_{j}-\varphi_{i}}{2}\right)} \, ,
 \label{eq:arclength}
\end{equation}
which uses the latitude and longitude coordinates of the ray $i$ and of the gas particle $j$ that are defined in the reference frame positioned at the stellar particle; 

\item  For each ray, we find the gas particle for which the arc length $\Omega_{ij}$ is minimum;

\item  If the stellar particle does $N_{\rm inj}$ energy injections, where $N_{\rm inj}\leq N_{\rm rays}$, we randomly choose $N_{\rm inj}$ rays out of $N_{\rm rays}$ rays and inject the energy into the corresponding gas particles\footnote{Since a gas particle is not forbidden to have multiple rays with which it was found to have the smallest arc length, the number of particles in which the energy is injected is always less than or equal to min($N_{\rm inj}$, $N_{\rm rays}$). Generally, if a gas particle receives $N$ rays in a given time-step, where $0 <N \leq N_{\rm rays}$, the total SN energy injected into this particle is $N\, \Delta E_{\rm inj}$. This property holds only for the isotropic model; in the other neighbour-selection models, a gas particle can only be bound to a single ray.}. If $N_{\rm inj}> N_{\rm rays}$, we increase $\Delta E_{\rm inj}$ by $N_{\rm inj}/N_{\rm rays}$ and inject the particles corresponding to all $N_{\rm rays}$ rays with the updated value of $\Delta E_{\rm inj}$.
\end{enumerate}

\subsection{Minimum distance method}

Another plausible neighbour-selection model to consider is the \textit{minimum distance} model (henceforth, \textsf{min\_distance}), motivated by the idea that SN feedback needs to be as local as possible to the star. In this prescription, we sort gas neighbours according to their separations from the stellar particle. If a stellar particle has $N_{\rm inj}$ energy-injection events, then the $N_{\rm inj}$ closest gas particles will partake in the SN feedback. 

Similar to the isotropic selection, the selection of the closest neighbour for SN feedback is expected to reduce the density bias seen in the \textsf{mass-weighted} method, albeit not always. If $N_{\rm inj}$ and $N_{\rm rays}$ are comparable to (or greater than) the number of particles in the stellar kernel, the \textsf{min\_distance} method will turn into the \textsf{mass-weighted} method, while the \textsf{isotropic} method will remain isotropic regardless of the value of $N_{\rm inj}$.

\subsection{Minimum and maximum density methods}

Finally, we will examine two extreme models: \textit{minimum} and \textit{maximum density} (from here on, \textsf{min\_density} and \textsf{max\_density}). Although they are not physically motivated, in our work they will serve as an estimate of the maximum possible variation in galaxy properties due to the neighbour selection strategy. The implementation of these two density-based models is the same as in the \textsf{min\_distance} model but in place of distance we sort gas neighbours according to their densities. In the \textsf{min\_density} (\textsf{max\_density}) method, if a stellar particle has $N_{\rm inj}$ energy-injection events, then the $N_{\rm inj}$ least (most) dense gas particles will receive the energy.

The five neighbour-selection algorithms that we have introduced in this section are illustrated in Figure \ref{fig:feedback_models}. There we show five possible realisations of distributing four energy injections among the gas particles in the stellar kernel -- one realisation for each neighbour-selection model. We note that in our simulations the probability of having a number of energy injections greater than one in a given time-step is much smaller than unity (see $\S$\ref{subsec:stochastic_thermal} for more details), and in Figure \ref{fig:feedback_models} the value of four is chosen only to highlight the differences between the models.

\section{Numerical simulations}
\label{sec: numerical simulations}

\subsection{Code and setup}

To test the five neighbour-selection models described in $\S$\ref{sec: neighbour_selection}, we use the smoothed particle hydrodynamics astrophysical code \textsc{swift}\footnote{\textsc{swift} is publicly available at \url{http://www.swiftsim.com}.} \citep{2016arXiv160602738S,2018ascl.soft05020S}, using the density-energy SPH scheme \textsc{Sphenix} \citep{2021MNRAS.tmp.2912B} to solve the hydrodynamical equations. The $\textsc{sphenix}$ scheme is designed for cosmological simulations and has been demonstrated to perform well on various hydrodynamical tests on different scales \citep{2021MNRAS.tmp.2912B}. We use the same SPH parameter values as in \citet{2021MNRAS.tmp.2912B}, including the quartic spline for the SPH kernel and the Courant–Friedrichs–Lewy parameter $C_{\rm CFL}=0.2$, which sets the time-steps of gas particles. Furthermore, we do not allow the ratio between time-steps of any two neighbouring gas particles to be greater than $4$. The target SPH smoothing length in our simulations is set to $1.2348$ times the local inter-particle separation, which corresponds to an expected weighted number of gas neighbours in the kernel $N_{\rm ngb} = 64.91$ for a quartic-spline kernel.  

\subsection{Initial conditions}

\subsubsection{Isolated disk galaxy}

The initial conditions for the distributions of gas and stars in the galaxy are taken from \cite{springel2005}. Our model for a Milky Way-mass galaxy comprises a dark matter halo with a  \citet{hernquist1990} potential, a total mass $M_{200} = 1.37 \times 10^{12} \, \rm  M_{\, \odot}$, concentration $c=9.0$ and spin parameter $\lambda=0.033$. The dark-matter halo is implemented as an analytic potential (for details see Nobels et al., in preparation) with virial velocity $v_{200}$ and radius $R_{200}$ equal to $163$ km s$^{-1}$ and $223$ kpc, respectively. In this halo, we place an exponential disk of stars and gas with the total mass of $M_{\rm disk} = 0.04 \, M_{200} = 5.48 \times 10^{10} \, \rm  M_{\, \odot}$. The initial gas fraction in the disk is set to $30$ per cent and the gas initially has solar metallicity, $Z_\odot=0.0134$ \citep{2009ARA&A..47..481A}.  The vertical distribution of the stellar disk has a constant scale height, which is equal to $10$ per cent of the radial scale length.

The gas particle mass at our fiducial resolution is $m_{\rm gas} = 10^{5}$ $\rm M_{\odot}$, corresponding to a Plummer-equivalent gravitational softening length of $0.2$ kpc. To investigate the dependence of our results on resolution, we also ran simulations of the same galaxy at resolutions of $m_{\rm gas} = 1.25 \times 10^{4}$ $\rm M_{\odot}$ and $m_{\rm gas} = 8 \times 10^{5}$ $\rm M_{\odot}$. The former corresponds to a softening of $0.1$ kpc, and the latter to $0.4$ kpc.

In order to increase the stability of the disk in the first $\approx 0.1$ Gyr of the simulations, we assigned a distribution of stellar ages to the stellar particles present in the initial conditions assuming a constant star formation rate of $10\, \rm M_{\rm \odot}$ yr$^{-1}$, which is a typical value for a young, star-forming Milky Way-mass galaxy. In this process, we only considered stellar particles whose cylindrical radius (in the disk plane) is less than $10$ kpc from the disk centre. 

\begin{table*}
\caption{Numerical simulations used in this work. Column (1) contains the names of the simulations; (2) $N_{\rm part}$ is the number of SPH particles in the simulation (gas + stars); (3) $m_{\rm gas}$ is the initial gas-particle mass; (4) $m_{\rm dm}$ is the mass of dark-matter particles (only in cosmological simulations); (5) $\varepsilon_{\rm soft,gas}$ is the Plummer-equivalent gravitational softening length for baryons; column (6) indicates whether the simulation uses an equation of state (i.e. an effective pressure floor); column (7) shows the algorithm for neighbour selection in SN feedback used in the simulation.}
\centering
\begin{tabular}{lrrlrll} 
Name & $N_{\rm part}$ & $m_{\rm gas}$ [$\rm M_\odot $] & $m_{\rm dm}$ [$\rm M_\odot $] & $\varepsilon_{\rm soft,gas}$ & EOS & SN feedback \\
	 
\hline
\hline
	    
\textsf{IG\_M5\_isotropic} & $82^3$ & $10^5$ & -- & $0.2$ pkpc & No & Isotropic \\
\textsf{IG\_M5\_mass\_weighted} & $82^3$ & $10^5$ & -- & $0.2$ pkpc &  No & Mass-weighted \\
\textsf{IG\_M5\_min\_distance} & $82^3$ & $10^5$ & -- & $0.2$ pkpc &  No & Minimum distance \\
\textsf{IG\_M5\_min\_density} & $82^3$ & $10^5$ & -- & $0.2$ pkpc &  No & Minimum density \\
\textsf{IG\_M5\_max\_density} & $82^3$ & $10^5$ & -- & $0.2$ pkpc &  No & Maximum density \\
\hline
\textsf{IG\_M5\_isotropic\_eos} & $82^3$ & $10^5$ & -- & $0.2$ pkpc & Yes & Isotropic \\
\textsf{IG\_M5\_mass\_weighted\_eos} & $82^3$ & $10^5$ & -- & $0.2$ pkpc & Yes & Mass-weighted \\
\textsf{IG\_M6\_isotropic} & $41^3$ & $8 \times 10^5$ & -- & $0.4$ pkpc &  No & Isotropic \\
\textsf{IG\_M6\_mass\_weighted} & $41^3$ & $8 \times 10^5$ & -- & $0.4$ pkpc &  No & Mass-weighted \\
\textsf{IG\_M4\_isotropic} & $164^3$ & $1.25 \times 10^4$ & -- & $0.1$ pkpc & No & Isotropic \\
\textsf{IG\_M4\_mass\_weighted} & $164^3$ & $1.25 \times  10^4$ & -- & $0.1$ pkpc &  No & Mass-weighted \\

\hline
\hline
	    
\textsf{COS\_M5\_isotropic} & $188^3$ & $2.26 \times 10^5$ &  $1.21 \times 10^6$  & min(0.89 ckpc, 0.35 pkpc) &  No & Isotropic \\
\textsf{COS\_M5\_mass\_weighted} & $188^3$ & $2.26 \times 10^5$ & $1.21 \times 10^6$  & min(0.89 ckpc, 0.35 pkpc)& No & Mass-weighted \\
\textsf{COS\_M5\_min\_distance} & $188^3$ & $2.26 \times 10^5$ & $1.21 \times 10^6$  & min(0.89 ckpc, 0.35 pkpc) &  No & Minimum distance \\
\textsf{COS\_M5\_min\_density} & $188^3$ & $2.26 \times 10^5$ & $1.21 \times 10^6$  & min(0.89 ckpc, 0.35 pkpc) &  No & Minimum density \\
\textsf{COS\_M5\_max\_density} & $188^3$ & $2.26 \times 10^5$ & $1.21 \times 10^6$  & min(0.89 ckpc, 0.35 pkpc) &  No & Maximum density \\
\end{tabular} \\

\label{tab:runs}
\end{table*}

\subsubsection{Cosmological simulations}

We ran a set of cosmological simulations in a periodic comoving volume of size ($6.25$ comoving Mpc)$^{3}$. The initial phases for this volume were taken from the public multiscale Gaussian white noise field \textit{Panphasia} \citep{2013MNRAS.434.2094J} and are described in \citet{2015MNRAS.446..521S} in Table B1. All cosmological simulations were started at redshift $z=127$. We use the \textit{Planck-13} cosmology \citep{Planck2013}:  $(\Omega_{\rm m,0}, \Omega_{\rm \Lambda,0}, \Omega_{\rm b,0}, h, \sigma_8, n_{\rm s}) =  (0.307, 0.693, 0.04825, 0.6777, 0.8288, 0.9611)$, where $\Omega_{\rm m,0}$,  $\Omega_{\rm b,0}$, and $\Omega_{\rm \Lambda,0}$ are the current density parameters of matter, baryons, and the dark energy, respectively; $h$ is the dimensionless Hubble constant; $\sigma_8$ is a linear $z=0$ rms value of Gaussian density fluctuations within $8$ $h^{-1}$Mpc spheres; and $n_s$ is the spectral index of primordial fluctuations.

The initial numbers of gas and dark matter particles are both equal to $188^3$. The gas-particle mass is $m_{\rm gas} = 2.26 \times 10^5 \, \rm M_\odot$ and the dark-matter particle mass $m_{\rm dm} = (\Omega_{\rm m,0}/\Omega_{\rm b,0}-1) \, m_{\rm gas} = 1.21 \times 10^6 \, \rm M_\odot$. The Plummer-equivalent gravitational softening length of baryons was set to the minimum of $0.89$ comoving kpc and $0.35$ proper kpc. The softening of dark matter was set to that of the baryons multiplied by $(\Omega_{\rm m,0}/\Omega_{\rm b,0})^{1/3}$. The simulations do not include supermassive black holes.

In this work, we only present results at $z=0$. To identify haloes in the $z=0$ snapshots, we used the publicly available structure finder \textsc{VELOCIraptor}\footnote{\href{https://velociraptor-stf.readthedocs.io/en/latest/}{https://velociraptor-stf.readthedocs.io/en/latest/}} \citep{2019MNRAS.482.2039C,Elahi2019}.

\subsection{Subgrid model for galaxy evolution}

\subsubsection{Cooling and heating}

Radiative gas cooling and heating are modelled using pre-computed tables from \cite{2020MNRAS.497.4857P}\footnote{We use the fiducial version the cooling tables, \textsf{UVB\_dust1\_CR1 \_G1\_shield} (for the naming convention and more details we refer the reader to Table 5 in \citealt{2020MNRAS.497.4857P}). } generated with the photoionization code \textsc{cloudy} \citep{Cloudy17}. The model of \cite{2020MNRAS.497.4857P} assumes the gas to be in ionization equilibrium in the presence of a modified version of the redshift-dependent, ultraviolet/X-ray background of \cite{2020MNRAS.493.1614F}, cosmic rays, and a local interstellar radiation field. The interstellar radiation field, the dust-to-metals ratio and the shielding length all depend on the density and temperature of the gas.

In our fiducial model for the ISM in cosmological and isolated galaxy simulations, gas particles are allowed to cool down to temperatures as low as $10$ K. Additionally, as one of the variations, we consider a case with a constant Jeans mass pressure floor as in \citet{2008MNRAS.383.1210S}, $P_{\rm eos} \propto \rho_{\rm gas}^{4/3}$, normalised to temperature  $T= 8 \times 10^3 \, \mathrm{K}$ at density $n_{\rm H} = 0.1 \, \mathrm{cm}^{-3}$. We include this variation to show that our results are not a mere consequence of the very high density gradients caused by the lack of pressure floor, and because it is common practice to include a pressure floor in simulations of galaxy formation at these resolutions.

\subsubsection{Star formation}

To decide whether a gas particle is star-forming or not, we use a temperature-density criterion. The cold ($T \ll 10^{4}$ K ) interstellar gas phase is expected to be unstable to star formation \citep[e.g.][]{Schaye2004}. Therefore, we allow gas particles to be star-forming if their temperature $T< 10^3$ K or their density\footnote{In the runs with an effective pressure floor, we use the subgrid temperature to decide whether a gas particle is star forming. The subgrid temperature is computed assuming thermal and pressure equilibrium between the unresolved cold gas phase and the effective pressure given by the floor (for details see Ploeckinger et al., in preparation).}, expressed in units of hydrogen particles per cubic cm, $n_{\rm H}$, exceeds $10^{2} \, \rm cm^{-3}$. The latter condition is essential for cosmological simulations at high redshifts where low-metallicity gas may not able to cool below $\sim 10^{4}$ K at resolved densities.

In our model, star formation occurs stochastically: if a gas particle satisfies the temperature-density criterion, its star formation rate, $\dot m_{\rm sf}$, is computed following the \citet{1959ApJ...129..243S} law 
\begin{equation}
\dot m_{\rm sf} = \varepsilon \frac{m_{\rm gas}}{t_{\rm ff}}\, ,
\end{equation}
where $t_{\rm ff} = [3\uppi/(32 G \rho)]^{1/2}$ is the free-fall time and $\varepsilon = 0.01$ is the efficiency of star formation per free fall time. We then compute the probability of this gas particle turning into a stellar particle by multiplying $\dot m_{\rm sf}$ by the particle's current time-step and dividing by its mass $m_{\rm gas}$. 

\subsubsection{SN feedback}
\label{subsec:stochastic_thermal}

For SN (type-II) feedback, we adopt the thermal stochastic subgrid model of \citetalias{2012MNRAS.426..140D} and use it in combination with the original, \textsf{mass-weighted} method or one of the four new neighbour-selection methods from $\S$\ref{sec: neighbour_selection}.

In simulations of galaxies with mass resolution as in our work, stellar particles represent populations of stars with a certain age, metallicity and initial mass function (IMF) $\Phi(m)$. Given the stellar IMF, the total number of stars ending their lives as core-collapse SNe per unit stellar mass can be computed as

\begin{equation}
    n_{\rm SN} = \int_{m_{\rm min}}^{m_{\rm max}} \, \Phi(m) \, \mathrm{d} m \, ,
\end{equation}
where $m_{\rm min}$ and $m_{\rm max}$ are the minimum and maximum mass of stars that die as core-collapse SNe, respectively. For the \citet{Chabrier2003} IMF with the lower and upper mass limits of $m_{\rm min}=8 \, \rm M_\odot$  and $m_{\rm max}=100 \, \rm M_\odot$, this gives $n_{\rm SN}=1.18 \times 10^{-2} \, \rm M^{-1}_\odot$. The SN energy budget per stellar particle of mass $m_{\rm *}$ can thus be written as
\begin{equation}
   \frac{ E_{\rm SN,tot} }{10^{51} \, \mathrm{erg} } = n_{\rm SN} \, f_{\rm E} \, \, m_{\rm *} = f_{\rm E} \, 1.18 \, \times 10^{2} \, \left(\frac{m_{\rm *}}{10^{4} \, \mathrm{M_{\odot}}}\right) \, ,
\end{equation}
where the free parameter $f_{\rm E}$ controls how much SN energy is liberated per single SN, in units of $10^{51}$ erg. \citetalias{2012MNRAS.426..140D} showed that in order to minimize the overcooling losses in SN thermal feedback, gas elements need to be heated to a temperature $T$ for which the cooling time-scale $t_{\rm c}$ (of the heated gas element) is roughly a factor of $10$ greater than the sound-crossing time-scale $t_{\rm s}$ (across the heated element). More precisely, they showed that assuming (i) the gas cooling rate is dominated by Brehmsstrahlung (ii) and the gas is fully ionized and has a primordial composition with the hydrogen mass fraction $X=0.752$, the ratio $f_{\rm t} \equiv t_{\rm c} / t_{\rm s}$ can be written in the form

\begin{equation}
    f_{\rm t} \equiv \frac{t_{\rm c}}{t_{\rm s}} = 68 \left(\frac{n_{\rm H}}{1 \, \rm cm^{-3}}\right)^{-2/3} \left(\frac{T}{10^{7.5} \, \rm K}\right)\left(\frac{m_{\rm ngb}}{10^{7}\, \rm M_\odot}\right)^{-1/3} \, ,
    \label{eq:ratio_time_scales}
\end{equation}
where $n_{\rm H}$ is the (hydrogen) number density of the local gas and $m_{\rm ngb}$ is the total mass in the gas elements neighbouring the stellar particle (if the star has $N_{\rm ngb}$ gas neighbours, all of which are of mass $m_{\rm gas}$, then $m_{\rm ngb} = N_{\rm ngb} \, m_{\rm gas}$). The energy corresponding to a temperature increase $\Delta T$ of one such element (a particle or cell) of mass $m_{\rm gas}$ is
\begin{align}
    \Delta E_{\rm inj}(m_{\rm gas},\Delta T) &= \frac{k_\mathrm{B}\Delta T}{(\gamma-1)}\frac{m_{\rm gas}}{\mu m_\mathrm{p}} \nonumber \\ 
    &= 1.3 \times 10^{53}\, \mathrm{erg}\, \left(\frac{m_{\rm gas}}{10^4 \,\rm M_{\odot}}\right)\left(\frac{\Delta T}{\mathrm{10^{7.5} \, K}}\right) \, ,
    \label{eq: heating_energy}
\end{align}
where $\gamma = 5/3$ is the ratio of specific heats for an ideal monatomic gas, $k_\mathrm{B}$ is the Boltzmann constant, $m_\mathrm{p}$ is the proton mass, and $\mu=0.6$ is the mean molecular weight of fully ionized gas.  If a stellar particle injects its neighbours with energy $\Delta E_{\rm inj}$ and all gas elements have mass $m_{\rm gas}$, then on average this stellar particle will have 
\begin{align}
\langle N_{\rm inj, tot} \rangle(m_*, f_{\rm E}, m_{\rm gas},\Delta T) &= \frac{E_{\rm SN, tot}(m_*, f_{\rm E})  \, }{\Delta E_{\rm inj}(m_{\rm gas},\Delta T)} \nonumber \\
&= 0.91 \, f_{\rm E} \,  \, \left(\frac{m_{\rm *}}{m_{\rm gas}}\right)\, \left(\frac{\Delta T}{\mathrm{10^{7.5} \, K}}\right)^{-1}  ,
\label{eq:N_heat}
\end{align}
energy injections over its lifetime\footnote{The numeral prefactor in the above equation is slightly lower than in \citetalias{2012MNRAS.426..140D} because we are using different integration limits for the stellar initial mass function: $m_{\rm min} = 8 \, \rm M_\odot$ instead of $6 \, \rm M_\odot$.}. Based on equations (\ref{eq:ratio_time_scales}) and (\ref{eq:N_heat}), \citetalias{2012MNRAS.426..140D} concluded that the optimal heating temperature should be around $\Delta T = 10^{7.5}$ K: making $\Delta T$ lower will decrease $f_{\rm t}$ resulting in a weaker SN feedback and stronger overcooling losses, while making $\Delta T$ significantly higher will cause undersampling of SNe feedback events (on average much less than one heating event per stellar particle). 

The probability that a star particle of age $t_j$ heats a particular gas neighbour in a time interval from $t_j$ to $t_j+\Delta t_{j}$ can be written as
\begin{equation}
    p_{\rm heat}(m_{\rm ngb},\Delta T, t_j, \Delta t_j) = \frac{\Delta E_{\mathrm{SN}, j}(t_j, \Delta t_j)}{\Delta E_{\rm inj}(m_{\rm ngb},\Delta T)} \, ,
    \label{eq:prob_heat}
\end{equation}
where $\Delta E_{\mathrm{SN},j}$ is the SN energy released during the time interval of length $\Delta t_j$ and $\Delta E_{\rm inj}(m_{\rm ngb},\Delta T)$ is the energy needed to heat the total gas mass in the stellar kernel, $m_{\rm ngb}$, by a temperature $\Delta T$. The energy $\Delta E_{\mathrm{SN}, j}$ is related to the total SN energy budget carried by the star particle $E_{\rm SN, tot}$ via

\begin{equation}
    E_{\rm SN, tot} = 10^{51} \, \mathrm{erg} \, f_{\rm E} \, \, m_{\rm *}   \,  \sum_{j} \, \int_{m_{\rm d}(t_{j}+\Delta t_{j})}^{m_{\rm d}(t_{j})}  \, \Phi(m) \, \mathrm{d}m \equiv \sum_{j} \, \Delta E_{\mathrm{SN},j}\, ,
    \label{eq: energy_per_dt}
\end{equation}
where the function $m_{\rm d}(t)$ gives the mass of the star(s) dying as core-collapse SNe at age $t$ and in this work is computed using the metallicity-dependent stellar-lifetime tables from \citet{Portinari1998}. The values of the function $m_{\rm d}(t)$ are limited to the range $m_{\rm min} \leq m_{\rm d}(t) < m_{\rm max}$, which is a consequence of the adopted IMF. As a result, the integral in the sum in equation (\ref{eq: energy_per_dt}) is non-zero only when the stellar age $t_j$ is roughly within $3 \, \mathrm{Myr} \leq t_j < 40 \, \mathrm{Myr}$, where the two numbers are the (rounded) lifetimes of stars with initial masses of $m_{\rm max}$ and $m_{\rm min}$, respectively. The number of terms in the sum in equation (\ref{eq: energy_per_dt}) depends on the size of star particles' time-steps.

\begin{figure*}
 \centering
  \includegraphics[width=0.999\textwidth]{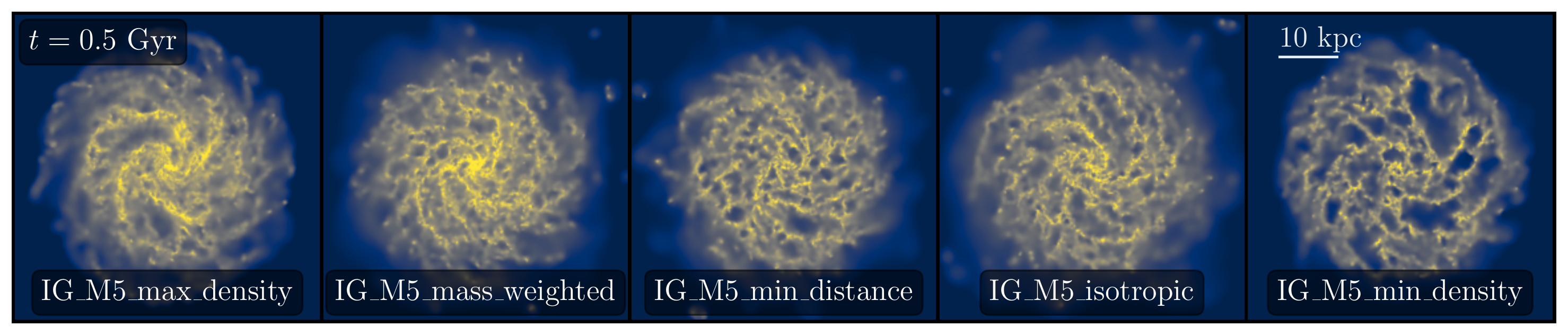} \\
    \includegraphics[width=0.999\textwidth]{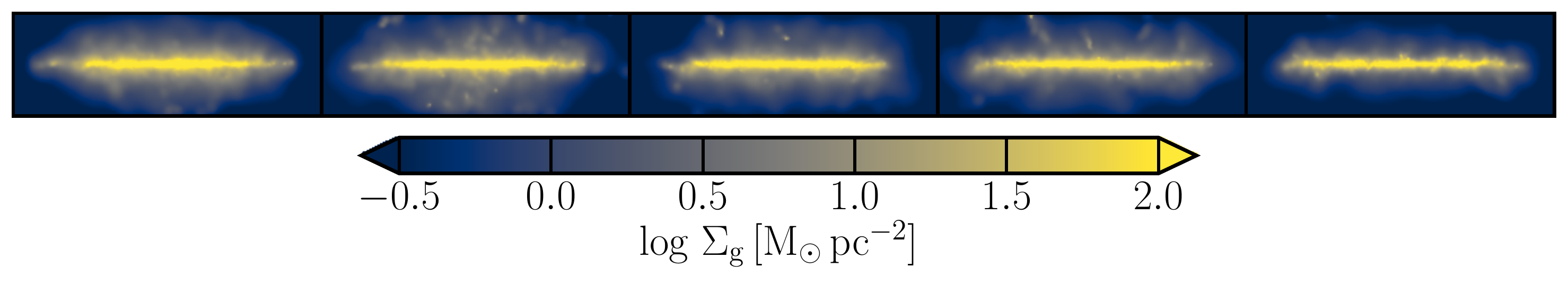}
    \caption{The distribution of gas in the isolated galaxy simulations with resolution $m_{\rm gas} = 10^{5} \, \rm M_\odot$ shown face-on (\textit{top}) and edge-on (\textit{bottom}) at time $t=0.5$ Gyr, for five methods of selecting gas particles neighbouring young stellar particles for SN feedback. The colour indicates gas surface density. The efficiency of SN feedback -- which differs solely due to the variations in the neighbour selection approach -- increases from left to right. In the runs with more efficient SN feedback, the gas is more dispersed and the gas surface densities are on average lower.}
    \label{fig:isolated_galaxy_morphology_gas_surface_density}
  \end{figure*}

Finally, to compute the number of thermal energy injection events $N_{\rm inj}$ in this time-step, the following algorithm is employed:

\begin{enumerate}
    \item For a given stellar particle with available energy $\Delta E_{\mathrm{SN}, j}$\footnote{Note that in place of sampling energy $\Delta E_{\mathrm{SN}, j}$ every time-step, \citetalias{2012MNRAS.426..140D} compute the heating probability just once using the total energy budged $E_{\rm SN, tot}$ when a stellar-particle's age becomes greater than $30$ Myr.} and $N_{\rm ngb}$ gas neighbours comprising a total mass $m_{\rm ngb}$, the heating probability $p_{\rm heat}$ is calculated following equation (\ref{eq:prob_heat})\footnote{If $p_{\rm heat}>1$, then the heating temperature $\Delta T$ is increased such that the new value of $p_{\rm heat}$ is equal to one.} and $N_{\rm inj}$ is initialised with $0$; 
    \item For each gas neighbour out of $N_{\rm ngb}$, a random number is drawn from a uniform distribution $0\leq r < 1$;
    \item If the random number is smaller than $p_{\rm heat}$, then the value of $N_{\rm inj}$ is incremented by one. 
\end{enumerate}

After having computed $N_{\rm inj}$, we pass it to the neighbour-selection algorithm used in the simulation and follow the remaining steps as described in $\S$\ref{sec: neighbour_selection} for the selected algorithm. We always use an SN energy fraction $f_{\rm E} = 2.0$ , which corresponds to $2 \times 10^{51}$ erg per SN, and the target heating temperature $\Delta T_{\rm SN} = 10^{7.5}$ K. According to equation (\ref{eq:N_heat}), with such parameters the expected number of thermal-injection events over the lifetime of a stellar particle is $\approx 1.8$ (and per time-step the expected number of events is $\ll 1$). We therefore use a maximum number of $N_{\rm rays} = 5$ rays per stellar particle in order to never run out of rays even in the most unlikely scenarios. By running additional tests (not presented here), we verified that as long as the average number of heating events per time step per particle remains below or similar to one, our results also remain valid for a factor-of-a-few higher and lower values of $f_{\rm E}$ and $\Delta T_{\rm SN}$.

Besides heating gas to high temperatures, massive stars also enrich their surroundings with metals. In our simulations, the enrichment is done continuously following \citet{2009MNRAS.399..574W} and \citet{2015MNRAS.446..521S}.

For simplicity, we do not include SN type-Ia-related processes (energy feedback and metal enrichment), which would have negligible impact on our results.

\subsubsection{Early stellar feedback}

We use \textsc{BPASS} \citep{BPASS2017,BPASS2018} version 2.2.1 with a \citet{Chabrier2003} initial mass function as stellar evolution model for all early feedback processes. The minimum and maximum stellar masses are $0.1 \, \rm M_\odot$ and $100 \, \rm M_\odot$, respectively. The early stellar-feedback processes we include in the simulations are stellar winds, radiation pressure and HII regions. 

The implementation and effects of these three early feedback processes will be described in detail in Ploeckinger et al. (in preparation). Briefly, stellar winds inject cumulative momentum following the \textsc{BPASS} tables. This wind feedback is implemented as stochastic kicks of gas particles with a kick velocity of $50$ km s$^{-1}$. Radiation pressure is implemented in the same way as stellar winds and is based on the \textsc{BPASS} photon energy spectrum and the optical depth computed following \citet{2020MNRAS.497.4857P}. Finally, young star particles stochastically ionize and heat neighbouring gas particles to $T = 10^{4}$ K where the probability of becoming an HII region is a function of the gas density and \textsc{BPASS} ionising photon flux. Gas particles tagged as HII regions are not allowed to form stars.

\subsection{Runs}

All runs presented in this work are summarised in Table \ref{tab:runs}. Names of isolated galaxy runs and of cosmological runs begin with a prefix \textsf{IG} and \textsf{COS}, respectively. The name also indicates which neighbour selection model (from $\S$\ref{sec: neighbour_selection}) is used: \textsf{isotropic}, \textsf{min\_distance}, \textsf{mass-weighted}, \textsf{min\_density}, or \textsf{max\_density}.  If a run uses an effective pressure floor, we add the suffix \textsf{\_EOS} to its name. The resolution of the isolated galaxy runs is indicated in the names via \textsf{\_M4}, \textsf{\_M5}, and  \textsf{\_M6} corresponding to the gas-particle mass of $m_{\rm gas} = 1.25 \times 10^{4} \, \rm M_\odot$, $m_{\rm gas} = 10^{5} \, \rm M_\odot$, and $m_{\rm gas} = 8 \times 10^{5} \, \rm M_\odot$, respectively. Finally, in the cosmological runs, \textsf{\_M5} stands for $m_{\rm gas} = 2.26 \times 10^{5} \, \rm M_\odot$, which is the only resolution we explore.

\section{Results}
\label{sec: Results}

\subsection{Isolated galaxy simulations}

We first show the results from the isolated galaxy simulations. Unless stated otherwise, all runs presented here have a particle mass of $m_{\rm gas} = 10^{5} \, \rm M_\odot$ and do not use an effective pressure floor. We mainly focus on the three neighbour-selection methods: \textsf{isotropic}, \textsf{min\_distance}, and \textsf{mass-weighted}. Additionally, we make use of the two more extreme methods, \textsf{min\_density} and \textsf{max\_density}. These two density-based methods together provide an estimate of the maximum variations in galaxy properties due to the neighbour selection for SN feedback.

\subsubsection{Morphology}

Figure \ref{fig:isolated_galaxy_morphology_gas_surface_density} shows the distribution of gas in the isolated galaxy simulations at time $t=0.5$ Gyr, colour-coded by gas surface density. The five galaxies seen in the plot were evolved with our five neighbour-selection methods for SN feedback (from left to right, \textsf{max\_density}, \textsf{mass-weighted}, \textsf{min\_distance}, \textsf{isotropic}, \textsf{min\_density}) and are shown face-on (top panels) and edge-on (bottom panels). The panels are arranged such that the runs with more efficient SN feedback (i.e. producing galaxies with less dense gas) -- solely due to the differences in the neighbour-selection algorithms -- are closer to the right of the plot. The neighbour-selection methods yielding more efficient SN feedback (\textsf{min\_distance}, 3rd panel; \textsf{isotropic}, 4th panel; \textsf{min\_density}, 5th panel) produce galaxies with more dispersed gas. In contrast, the gas in the galaxies evolved with less efficient methods (\textsf{max\_density}, 1st panel; \textsf{mass-weighted}, 2nd panel) is more concentrated, with the gas surface densities in the centre and within spiral arms typically exceeding $10^{2} \, \rm  M_{\odot} \, pc^{-2}$. 

\begin{figure}
 \centering
 \includegraphics[width=0.49\textwidth]{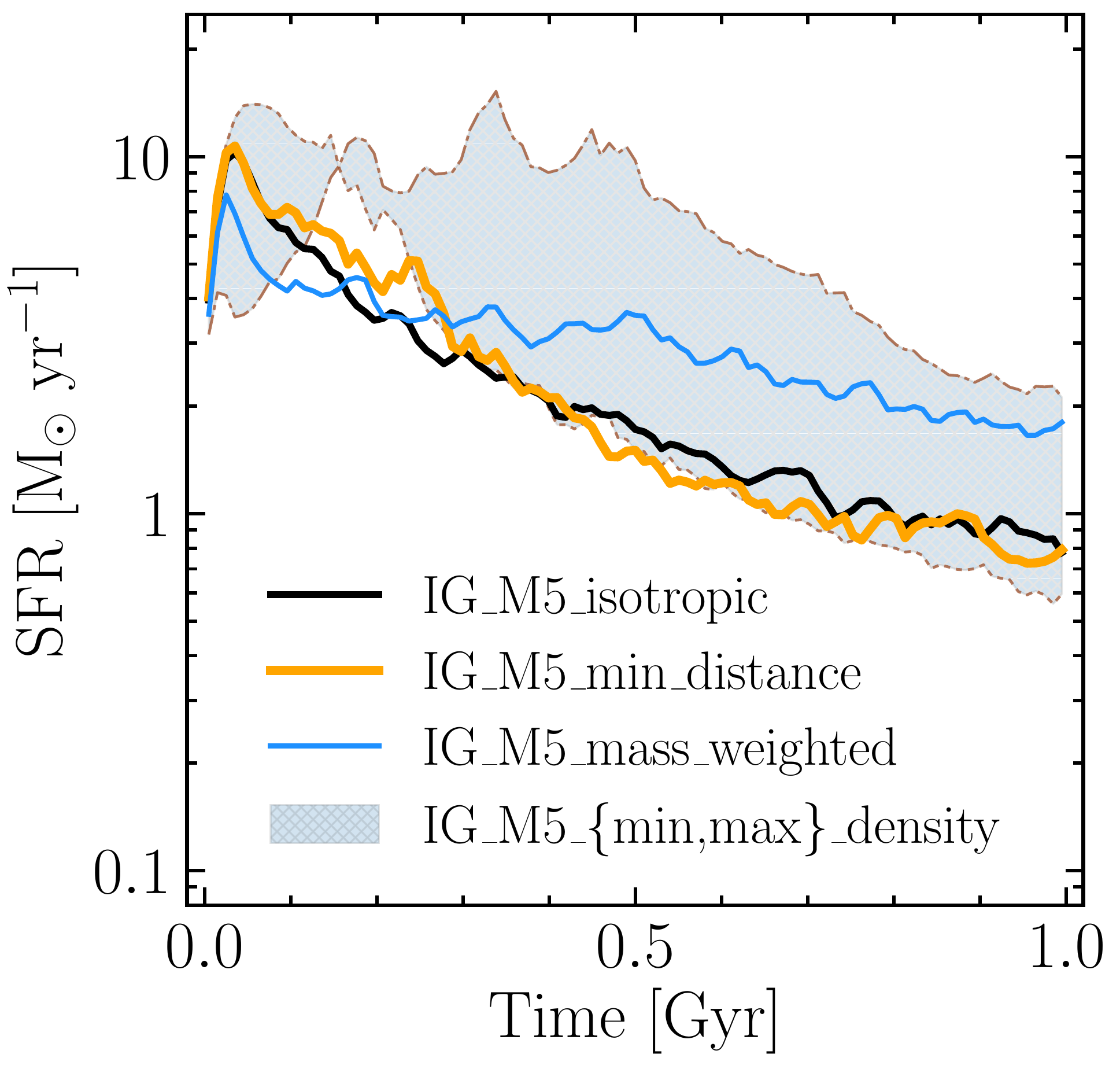}
 \caption{Star formation histories in the isolated galaxy simulations with resolution $m_{\rm gas} = 10^{5} \, \rm M_\odot$ with \textsf{isotropic} (\textit{black}), \textsf{min\_distance} (\textit{orange}) and \textsf{mass-weighted} (\textit{blue}) neighbour-selection methods for SN feedback. The hatched grey area provides an estimate of the maximum variation in the star formation rate due to the neighbour-selection strategy: its boundaries are defined by the star formation rates in the runs with the most and least efficient SN feedback: \textsf{IG\_M5\_min\_density} (\textit{dashed}) and \textsf{IG\_M5\_max\_density} (\textit{dash-dotted}), respectively.}
 \label{fig:star_formation_main}
 \end{figure}
 
 \subsubsection{Star formation rates}
 
Figure \ref{fig:star_formation_main} displays the star formation histories for the isolated galaxy simulations with the three main neighbour-selection methods: \textsf{isotropic} (black), \textsf{min\_distance} (orange), and \textsf{mass-weighted} (blue). For reference, we also show the runs with the two extreme models, \textsf{min\_density} and \textsf{max\_density}, shown as the upper and lower boundaries of the grey area. The star formation histories in \textsf{IG\_M5\_min\_density} and \textsf{IG\_M5\_max\_density} are further highlighted by the brown dashed and dash-dotted curves, respectively, to better distinguish the two runs. The star formation history in the \textsf{IG\_M5\_max\_density} run initially ($t\lesssim 0.15$ Gyr) defines the lower boundary of the grey area, but at all later times ($t\gtrsim 0.15$ Gyr) it corresponds to the upper boundary. The opposite is true for the \textsf{IG\_M5\_min\_density} run. 
 
All three main runs experience an initial burst of star formation, which ends at $t \approx 0.15$ Gyr. The star formation rates in the \textsf{isotropic} and \textsf{min\_distance} models are very much alike at all times and are systematically lower than in the \textsf{mass-weighted} model after the initial burst. At time $t = 0.5$ Gyr, the ratio of the star formation rate in the \textsf{mass-weighted} model to that in the \textsf{isotropic} (or \textsf{min\_distance}) is $\approx 2.0$, which increases to $\approx 2.5$ by $t = 1.0$ Gyr. The stars in the \textsf{IG\_M5\_mass\_weighted} run form at a higher rate because the \textsf{mass-weighted} method is biased towards heating gas at higher densities, where the injected SN energy dissipates faster, making the SN feedback less efficient at suppressing star formation.
 
\begin{figure}
 \centering
 \includegraphics[width=0.49\textwidth]{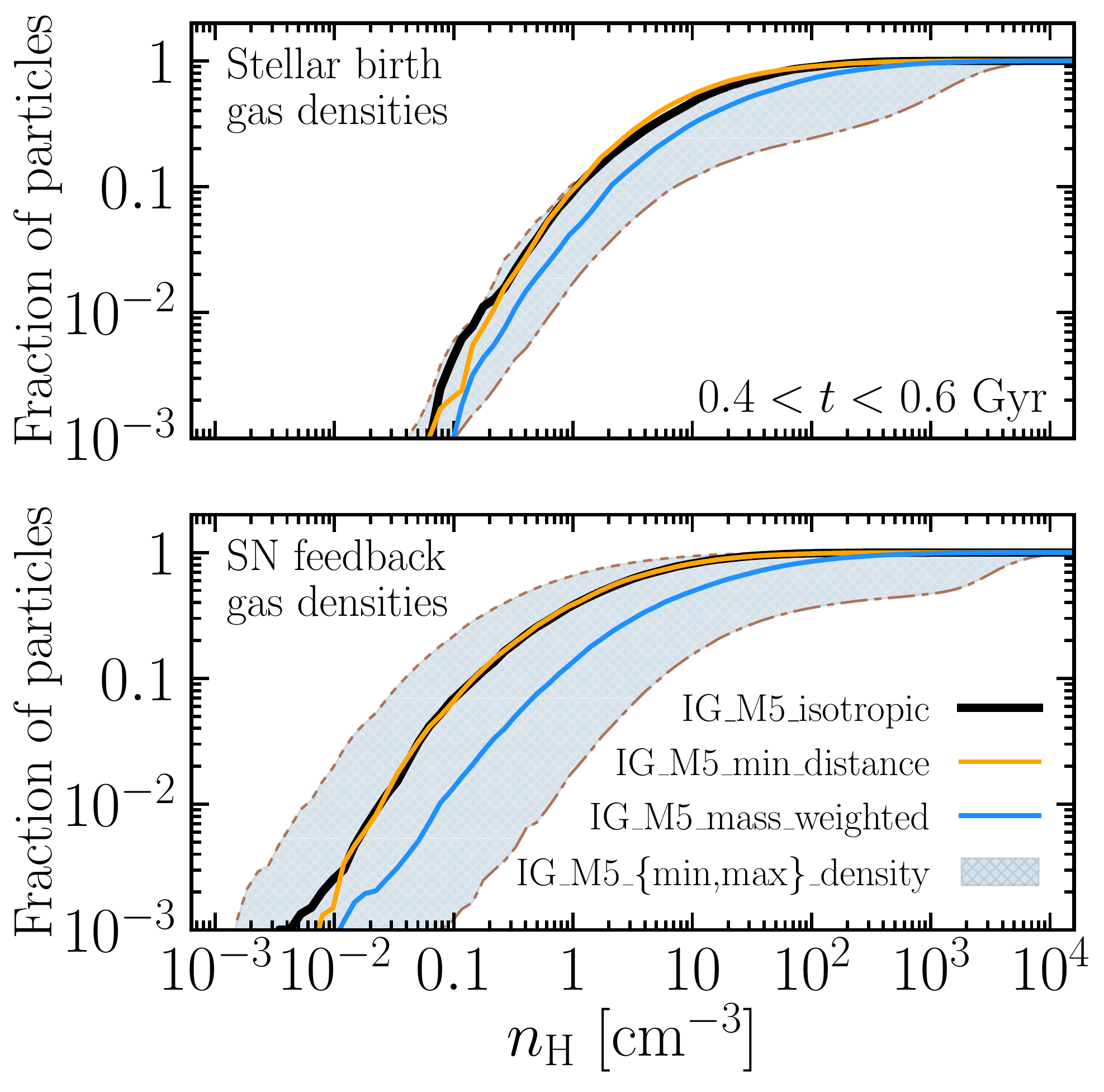}
 \caption{The cumulative distribution of stellar birth gas densities (\textit{top panel}) and SN feedback gas densities (\textit{bottom panel}) recorded at times $0.4 < t < 0.6$ Gyr, in the isolated galaxy simulations with resolution $m_{\rm gas} = 10^{5} \, \rm M_\odot$ with \textsf{isotropic} (\textit{black}), \textsf{min\_distance} (\textit{orange}) and \textsf{mass-weighted} (\textit{blue}) neighbour-selection methods for SN feedback. The hatched grey area is constructed in the same way as in Figure \ref{fig:star_formation_main} (i.e. using the \textsf{min\_density} and \textsf{max\_density} methods). Both stellar birth densities and SN feedback densities are sensitive to the method of neighbour selection.}
 \label{fig:density_main}
 \end{figure}
 
Comparing all five runs together, we find that star formation rates can differ by more than a factor of 5 depending on the method of neighbour selection. The \textsf{max\_density} model is initially the most effective at suppressing star formation. That is because it targets the \textit{densest} gas, where the majority of the stars form. This, however, only lasts while $t\lesssim 0.15$ Gyr after which the method becomes the least effective at preventing star formation. This is a consequence of the strongly enhanced radiative cooling losses in the densest gas where most of the SN energy is injected. The run \textsf{IG\_M5\_min\_density} has the opposite behaviour: SN feedback in this run is least inefficient at suppressing the initial star formation ($t\lesssim 0.15$ Gyr), because SNe heat the \textit{least} dense gas where not many stars form, but later it becomes most efficient owing to having the smallest radiative losses amongst the 5 considered runs.

 \subsubsection{Stellar birth densities and SN feedback densities}
 
 \begin{figure}
 \centering
 \includegraphics[width=0.49\textwidth]{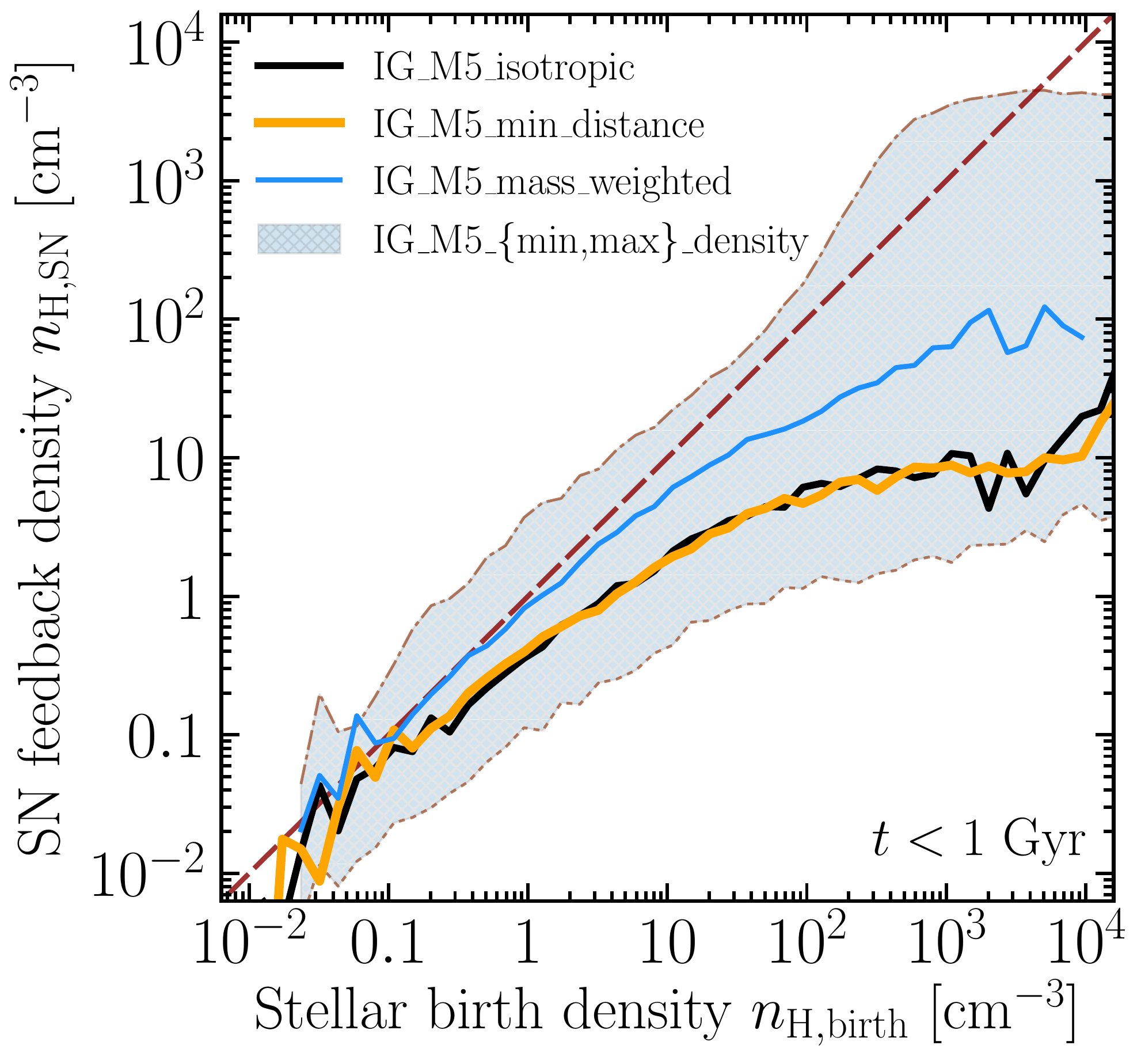}
 \caption{The median SN feedback densities versus stellar birth gas densities recorded at times $t<1$ Gyr, in the isolated galaxy simulations with resolution $m_{\rm gas} = 10^{5} \, \rm M_\odot$ with \textsf{isotropic} (\textit{black}), \textsf{min\_distance} (\textit{orange}) and \textsf{mass-weighted} (\textit{blue}) neighbour-selection methods. The hatched grey area is constructed using the \textsf{min\_density} and \textsf{max\_density} methods. The diagonal dashed line gives the $x=y$ relation. At a fixed stellar birth density, the median SN feedback density can vary by more than two orders of magnitude depending on the method of neighbour selection.}
 \label{fig:stellar_birth_density_stellar_feedback_density}
 \end{figure}

In Figure \ref{fig:density_main}, we display the cumulative distributions of stellar birth gas densities (top panel) and of SN feedback gas densities (bottom panel), which correspond, respectively, to the gas density at the time the star particle was formed and at the time the SN energy was injected. The stellar birth density distributions include all stars formed at times $0.4 < t < 0.6$ Gyr. For the SN gas density distributions, we used a tracer field carried by each gas particle in the simulation that would record the gas particle's density when it was last heated by SNe; we then collected the data from all gas particles that were heated by SNe at times $0.4 < t < 0.6$ Gyr. The curves again show the results for the three main neighbour-selection methods, while the shaded region for \textsf{min\_density} and \textsf{max\_density}. 
 
 \begin{figure}
     \centering
     \includegraphics[width=0.49\textwidth]{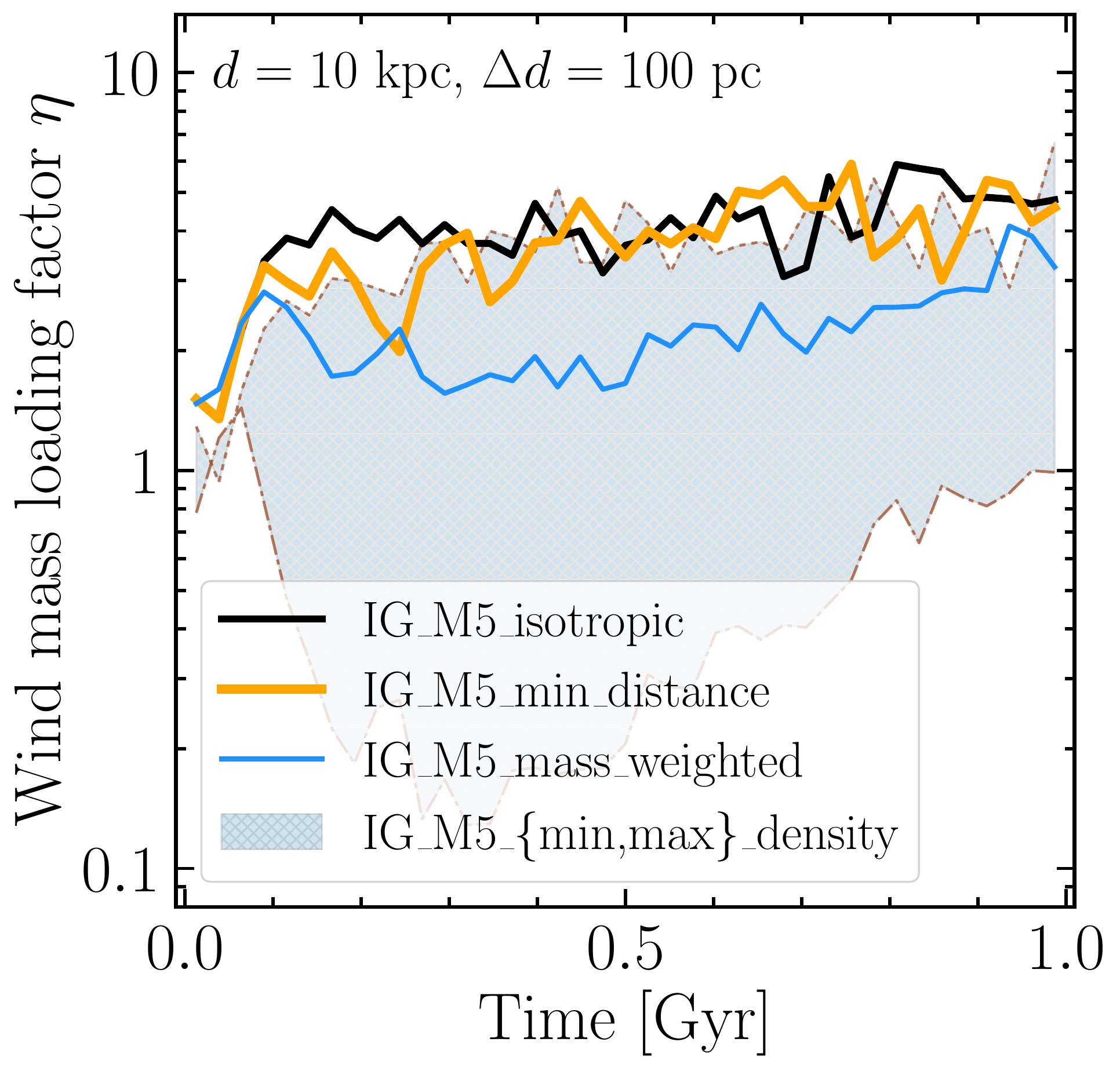}
     \caption{Gas wind mass loading factors at height $d=10\pm 0.1$ kpc above/below the galactic disk plotted versus time, in the isolated galaxy simulations with resolution $m_{\rm gas} = 10^{5} \, \rm M_\odot$ with \textsf{isotropic} (\textit{black}), \textsf{min\_distance} (\textit{orange}) and \textsf{mass-weighted} (\textit{blue}) neighbour-selection methods. The hatched grey area is constructed using the \textsf{min\_density} and \textsf{max\_density} methods. The highest mass loading factors are attained in the \textsf{min\_density}, \textsf{isotropic} and \textsf{min\_distance} models, followed by the intermediate mass loading factors in \textsf{mass-weighted}, and the lowest in \textsf{max\_density}.}
     \label{fig:mass_loadting_main}
 \end{figure}

 \begin{figure*}
 \centering
    \includegraphics[width=0.999\textwidth]{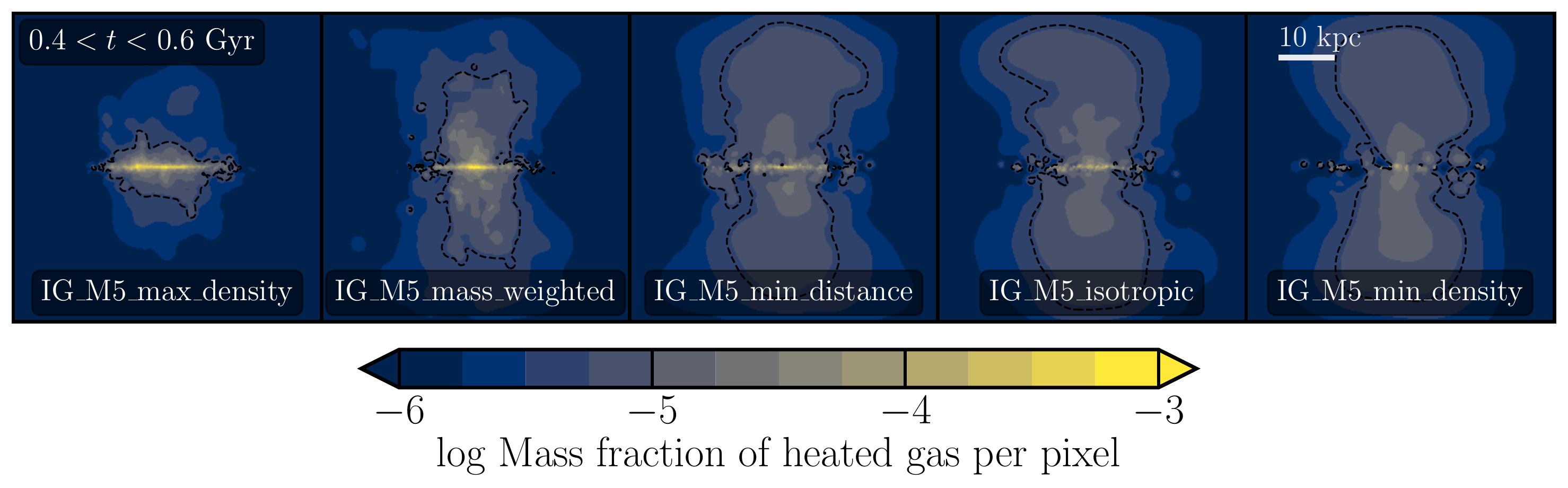}
    \caption{SN heated gas in the isolated galaxy simulations shown edge-on at $0.6$ Gyr. The colour scale indicates the mass fraction of the gas that was heated by SNe at times between $0.4$ and $0.6$ Gyr. Each panel corresponds to one of the five neighbour-selection methods for SN feedback. The black dashed contours are defined by the mass fraction of $5\times 10^{-6}$ and are shown to guide the eye. The efficiency of SN feedback and, hence, gas outflows -- which differs solely due to variations in neighbour selection -- increases from left to right.}
    \label{fig:mass_fraction_of_heated_gas}
 \end{figure*}
  
Both stellar birth densities and SN feedback gas densities are sensitive to the neighbour selection, though the latter varies more significantly. In the \textsf{isotropic} and \textsf{min\_distance} models, the gas turns into stars at nearly identical gas densities, which are lower than in the \textsf{mass-weighted} model by roughly a factor of 2. Similarly, the distributions of SN feedback gas densities look statistically identical for \textsf{isotropic} and \textsf{min\_distance} and the SN densities are lower than for \textsf{mass-weighted} by approximately a factor of 4. The differences between \textsf{min\_density} and \textsf{max\_density} are more dramatic: the stellar birth densities vary by more than an order of magnitude and the feedback densities by more than two orders of magnitude.
 
Next, in Figure \ref{fig:stellar_birth_density_stellar_feedback_density}, we show how median SN feedback densities are correlated with stellar birth densities. For this, we made each star particle record not only the gas density at which it was born, but also the density of the gas particle it heated in its most recent SN event. Here we consider all stellar particles that had at least one SN feedback event by the end of the simulation (1 Gyr).
 
At low gas densities ($n_{\rm H} \lesssim 1 \, \rm cm^{-3}$), the dynamical evolution of the gas is relatively slow and the density coherence length is relatively large, so that SNe go off in the ISM of roughly the same density as when the stars were formed. At higher gas densities ($n_{\rm H} \gtrsim 10 \, \rm cm^{-3}$), the situation is different: the relation between stellar birth and SN feedback densities begins to flatten, which happens because high-density clumps are short-lived and/or compact. In this case, stellar particles form in clusters, which means that before a given stellar particle has its first SN thermal injection, the parent gas cloud can already be dispersed by an SN blast originated from another stellar particle. Alternatively, a star particle may move out of its birth cloud before  injecting SN feedback, which will result in a lower gas density at the time of SN feedback.
 
The trends for the \textsf{isotropic}, \textsf{min\_distance}, and \textsf{mass-weighted} methods converge for $n_{\rm H,birth} \lesssim 0.1 \, \rm cm^{-3}$, which is a consequence of (i) little dynamical evolution of the gas between stellar-birth times and SN-feedback times, and (ii) the lack of steep density gradients at such low densities. However, at higher densities, the \textsf{mass-weighted} model begins to diverge from the \textsf{isotropic} and \textsf{min\_distance}, which themselves again show a high degree of resemblance. At the stellar birth density of $n_{\rm H,birth}=10^{3} \, \rm cm^{-3}$, the median SN density in the \textsf{isotropic} and \textsf{min\_distance} models is only $n_{\rm H,SN}\approx 10 \, \rm cm^{-3}$, whereas in the \textsf{mass-weighted} it is $n_{\rm H,SN}\approx 10^2 \, \rm cm^{-3}$. These SN densities are bracketed by the two more extreme models, which at $n_{\rm H,birth}=10^{3} \, \rm cm^{-3}$ have $n_{\rm H,SN} \approx 2  \, \rm cm^{-3}$ (\textsf{min\_density}) and $n_{\rm H,SN} \approx 3 \times 10^{3}  \, \rm cm^{-3}$ (\textsf{max\_density}). 

\begin{figure}
     \centering
     \includegraphics[width=0.49\textwidth]{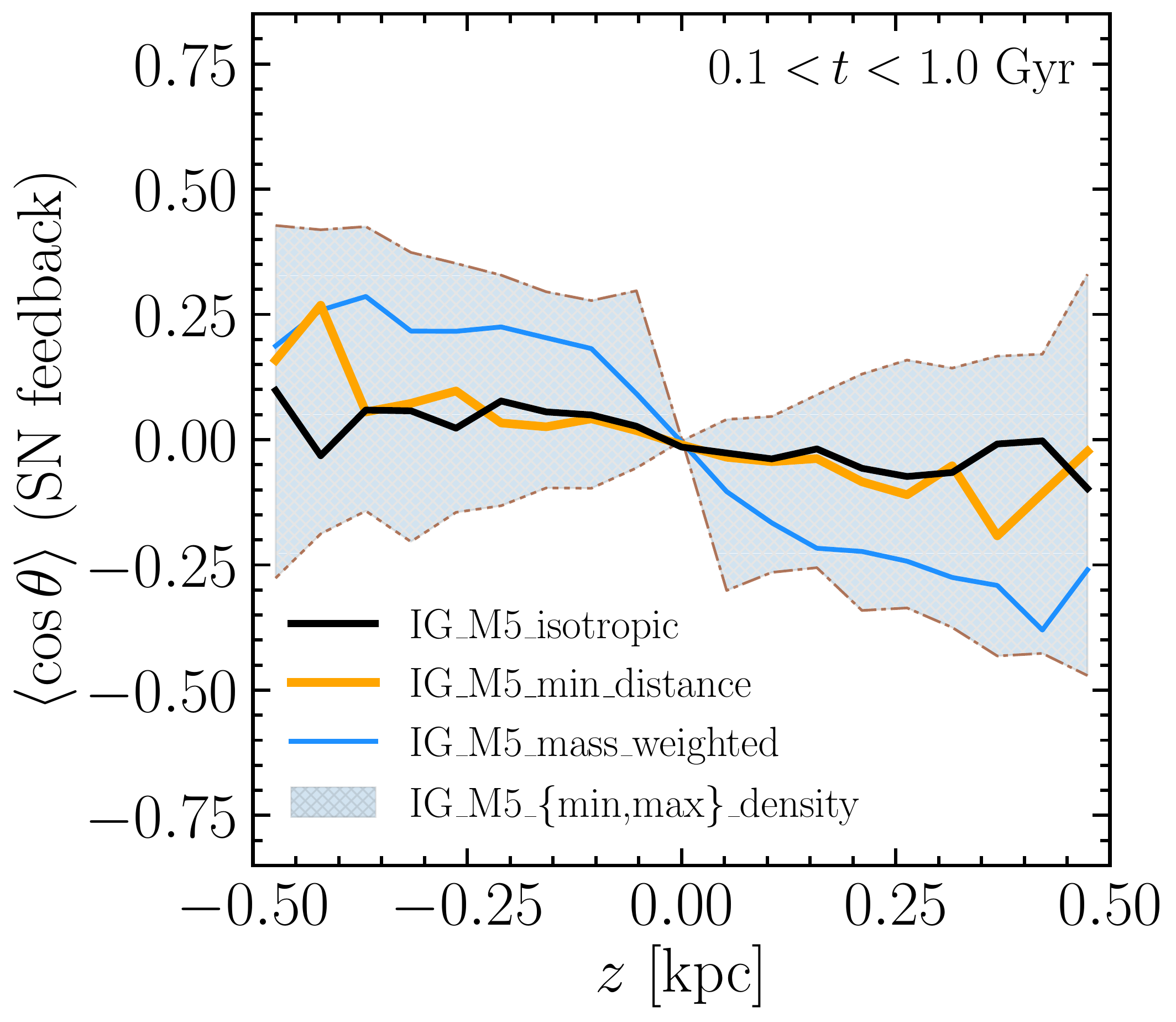}
     \caption{The average cosine of the angle between the vector from the star particle to the heated gas particle and the $z$ normal vector as a function of the height at which SN feedback occurred (see $\S$\ref{sec:isotropy} for details and equation  \ref{eq:aver_cosine} for the definition). If stars distributed the SN energy isotropically, then we would find $\langle \cos \theta \rangle = 0$. A negative slope of $\langle \cos \theta \rangle$ with $z$ indicates a bias towards the disk. The results are shown for the \textsf{isotropic} (\textit{black}), \textsf{min\_distance} (\textit{orange}) and \textsf{mass-weighted} (\textit{blue}) methods. The hatched grey area is constructed using \textsf{min\_density} and \textsf{max\_density}. Only the gas particles that were last heated by SNe between $t=0.1$ and $1.0$ Gyr are considered. Among the five methods, \textsf{isotropic} gives the most isotropic distribution of the azimuthal angles, as expected.}
     \label{fig:isotropy_main}
\end{figure}

\subsubsection{Wind mass loading factors}
  
Differences in the selection of gas neighbours also impact how effective SN feedback is at pushing gas out of the galaxy. To characterise the strength of gas outflows in isolated galaxies, we define the wind mass loading factor $\eta$ at time $t$ and at (absolute) height $d$ from the disk plane as 
\begin{equation}
    \eta(t, d, \Delta d) = \frac{1}{\dot{m}_{\rm sf, tot}} \sum_{|z_i\pm d|<\Delta d/2} \, \frac{m_{\mathrm{gas},i} \, |v_{z,i}|}{\Delta d}\, , 
\end{equation}
where $\dot{m}_{\rm sf, tot}$ is the galaxy total star formation rate at time $t$, $v_{z,i}$ is the velocity $z$ component of particle $i$, $m_{\mathrm{gas},i}$ is the mass of particle $i$, $z_i$ is the $z$ coordinate (height) of particle $i$ relative to the galactic disk, and the sum is computed over all outflowing (i.e. vertically 
moving away from the disk) gas particles whose heights are within $d\pm \Delta d /2$ from the disk (the disk plane extends in the $x$ and $y$ directions).

Figure \ref{fig:mass_loadting_main} shows the temporal evolution of the mass loading factor computed at height $d=10$ kpc in a height window $\Delta d = 0.1$ kpc, for the same runs as in the previous figures. It reveals that models with less (more) efficient SN feedback produce weaker (stronger) outflows, as expected. The \textsf{isotropic} (black curve) and \textsf{min\_distance} (orange curve) models have a more-or-less constant mass loading factor $\eta \approx 4$. In the \textsf{mass-weighted} model (blue curve), the mass loading is systematically lower, $\eta \approx 2$. Furthermore, we see that in the \textsf{isotropic} and \textsf{min\_distance} models, $\eta$ fluctuates around the grey area's upper boundary, which corresponds to the \textsf{IG\_M5\_min\_density} run. In other words, \textsf{isotropic} and \textsf{min\_distance} methods generate galactic winds with the mass loading that is on average as high as in the run with the most efficient SN feedback. Conversely, in \textsf{IG\_M5\_max\_density}, which defines the lower boundary of the grey area, SN feedback is so weak that $\eta \lesssim 1$ at all times.

To better understand the differences in $\eta$, we show in Figure \ref{fig:mass_fraction_of_heated_gas} the galaxy edge-on at time $0.6$ Gyr, colour-coded by the mass fraction of the gas heated by SNe between $0.4$ and $0.6$ Gyr. Each panel corresponds to one of the five neighbour-selection models. The panels are arranged such that the efficiency of SN feedback (and, hence, the strength of outflows) increases from left to right. This plot is consistent with Figure \ref{fig:mass_loadting_main}, confirming that in runs with less (more) efficient SN feedback, the gas is less (more) outflowing. The two extreme cases are a galactic fountain-like outflow, operating at distances $|z| \lesssim 10$ kpc from the disk, in the leftmost panel (\textsf{IG\_M5\_max\_density}); and the strong, stable wind, propagating beyond $|z| \approx 30$ kpc, in the rightmost panel (\textsf{IG\_M5\_min\_density}).

\subsubsection{Degree of isotropy of the SN feedback}
\label{sec:isotropy}

\begin{figure}
 \centering
 \includegraphics[width=0.49\textwidth]{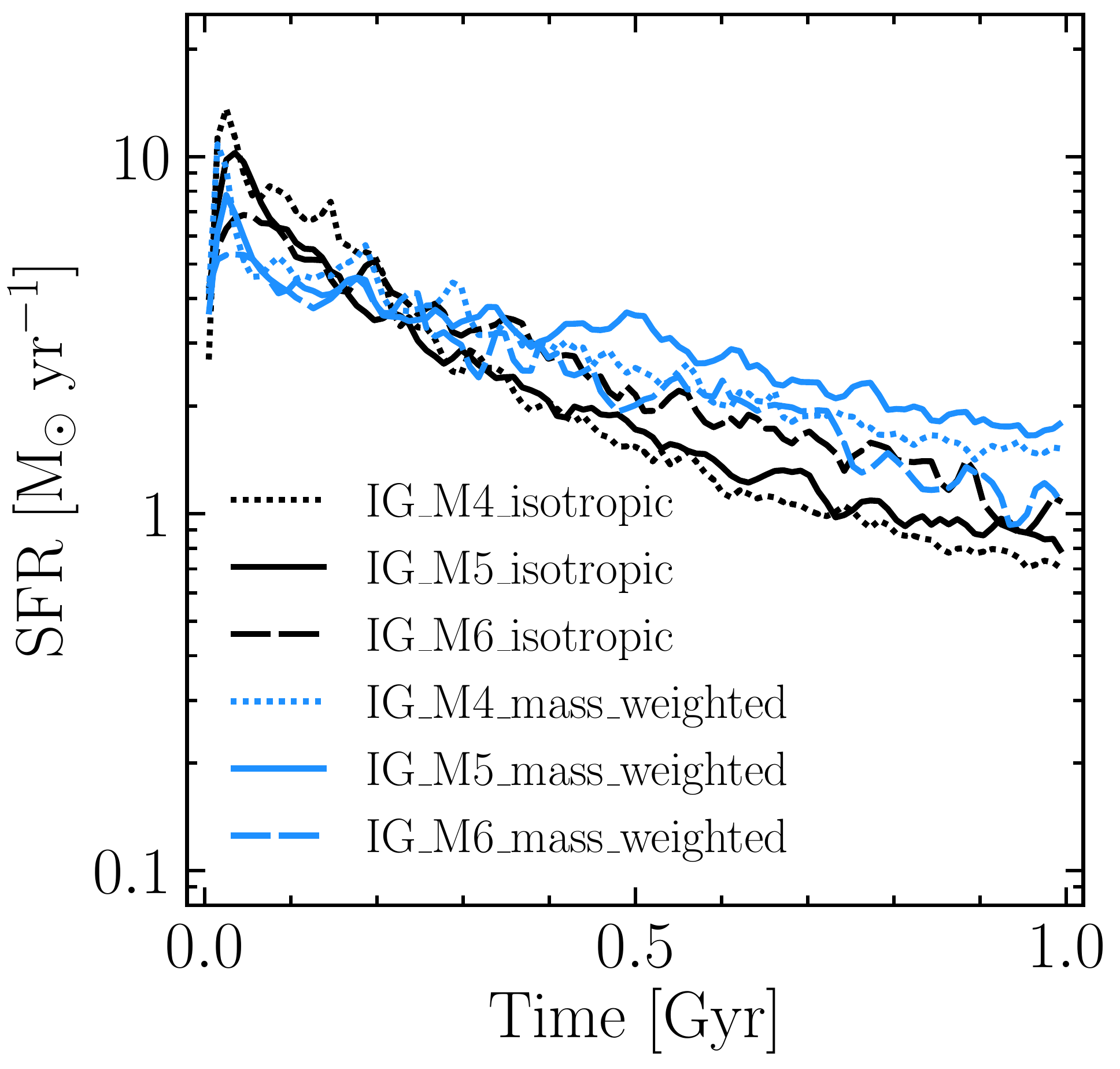}
 \caption{Star formation histories in the isolated galaxy simulations with the \textsf{isotropic} (\textit{black}) and \textsf{mass-weighted} (\textit{blue}) methods of neighbour selection shown at three resolutions: $m_{\rm gas} = 1.25 \times 10^{4} \, \rm M_\odot$ (M4, \textit{short-dashed}), $m_{\rm gas} = 10^{5} \, \rm M_\odot$ (M5, \textit{solid}), and $m_{\rm gas} = 8 \times 10^{5} \, \rm M_\odot$ (M6, \textit{long-dashed}). The differences in the star formation history between the \textsf{isotropic} and \textsf{mass-weighted} models increase with higher resolution. The \textsf{isotropic} and \textsf{mass-weighted} methods both show good convergence at our fiducial resolution M5, with the \textsf{isotropic} method being slightly better at times $t\gtrsim 0.5$ Gyr.}
 \label{fig:star_formation_res}
\end{figure}

\begin{figure}
 \centering
 \includegraphics[width=0.49\textwidth]{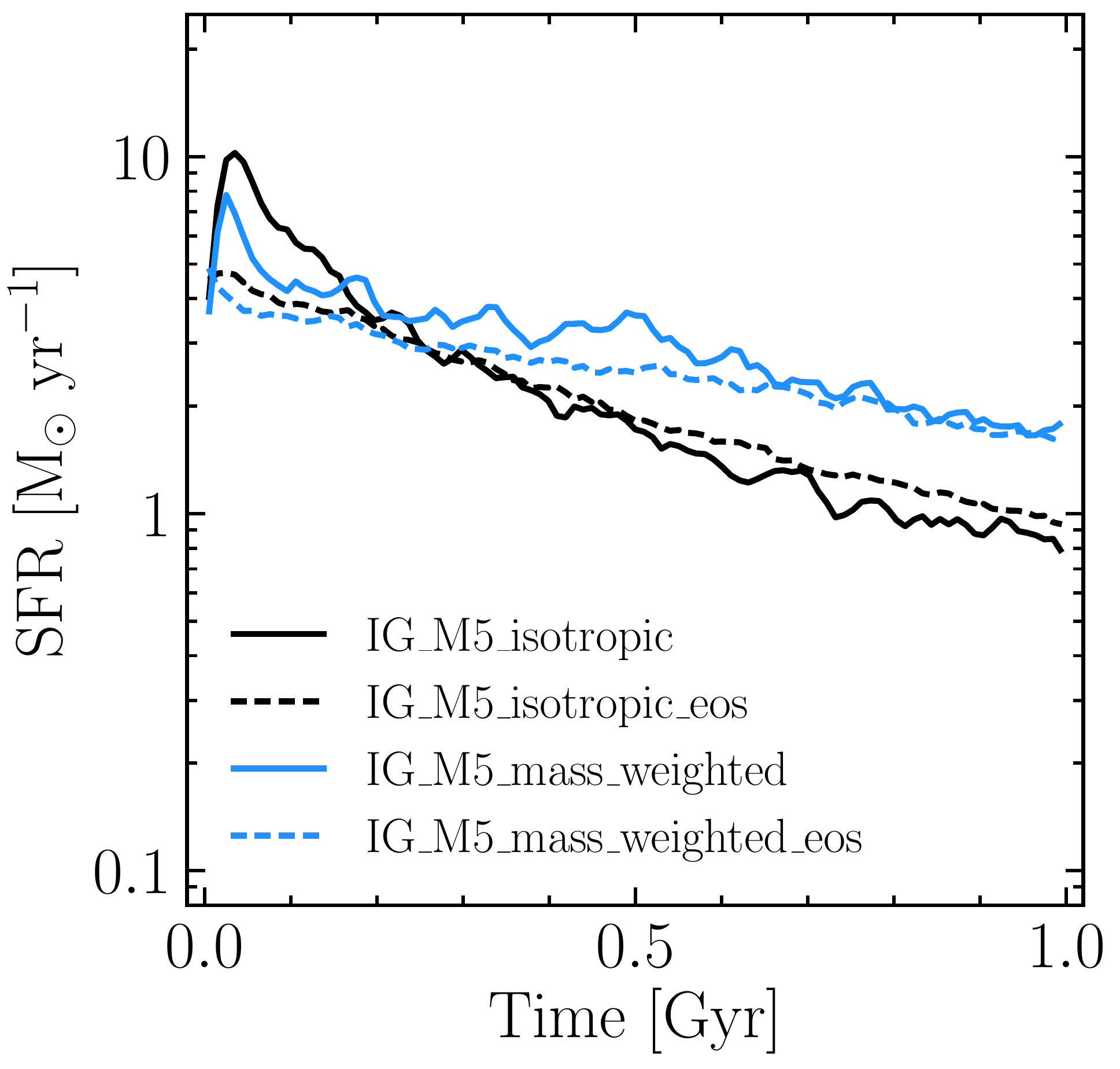}
 \caption{Star formation histories for the isolated galaxy simulations at resolution $m_{\rm gas} = 10^{5} \, \rm M_\odot$ with the \textsf{isotropic} (\textit{black}) and \textsf{mass-weighted} (\textit{blue}) methods, with (\textit{dashed}) and without (\textit{solid}) the effective pressure floor. While the inclusion of a pressure floor slightly decreases the differences, the effect of the neighbour selection algorithm remains similar.}
 \label{fig:star_formation_eos}
\end{figure}

Our next step is to characterise the isotropy of the SN feedback. Because the galaxy has an exponentially declining density along the $z$ direction, the density gradients in this direction are particularly steep. We are interested in knowing how much better our \textsf{isotropic} model performs in distributing SN energy isotropically under such highly inhomogeneous conditions, compared to the other neighbour-selection algorithms.

Figure \ref{fig:isotropy_main} shows the average cosine of the angle between the vector from the star particle to the heated gas particle and the $z$ normal vector, as a function of the height at which SN feedback events take place. More precisely, for a given $z-$coordinate bin with the two edges $z-\Delta z/2$ and $z+\Delta z/2$, the average cosine  is computed as

\begin{equation}
   \langle\cos \theta \rangle(z)= \frac{1}{N_{\rm tot}}\sum_{|\boldsymbol{r}_{\mathrm{star},i}\cdot \boldsymbol{n}_z -z|<\Delta z/2}\frac{(\boldsymbol{r}_{\mathrm{gas},i} - \boldsymbol{r}_{\mathrm{star},i}) \cdot \boldsymbol{n}_z}{ |\boldsymbol{r}_{\mathrm{gas},i} - \boldsymbol{r}_{\mathrm{star},i}|}\,  
   \label{eq:aver_cosine}
\end{equation}
where $N_{\rm tot}$ is the number of particles satisfying the condition $|\boldsymbol{r}_{\mathrm{star},i}\cdot \boldsymbol{n}_z -z|<\Delta z/2$, $\boldsymbol{r}_{\mathrm{gas},i}$ is the coordinate of the $i^{\rm th}$ gas particle at the time when it was heated by SNe\footnote{If the particle has never been heated by SN, it does not contribute to the sum.}, $\boldsymbol{r}_{\mathrm{star},i}$ is the coordinate of the stellar particle that heated the $i^{\rm th}$ gas particle, also recorded at the time of the SN feedback event; and $\boldsymbol{n}_z$ is the unit vector pointing in the $z$ direction, perpendicular to the disk plane.

The results are shown for the isolated galaxy simulations with resolution $m_{\rm gas} = 10^{5} \, \rm M_\odot$ with the \textsf{isotropic} (black), \textsf{min\_distance} (orange) and \textsf{mass-weighted} (blue) neighbour-selection algorithms; and the grey shaded region is again defined using the \textsf{IG\_M5\_min\_density} and \textsf{IG\_M5\_max\_density} runs. We only consider those gas particles that were last heated by SNe between $t=0.1$ Gyr and $1.0$ Gyr; we do not include the initial evolutionary stage ($t< 0.1$ Gyr) because the galaxy at such early times has a dearth of gas particles below and above the disk and because we have seen that this phase is strongly affected by the initial set-up.

\begin{figure*}
  \centering
  \includegraphics[width=0.995\textwidth]{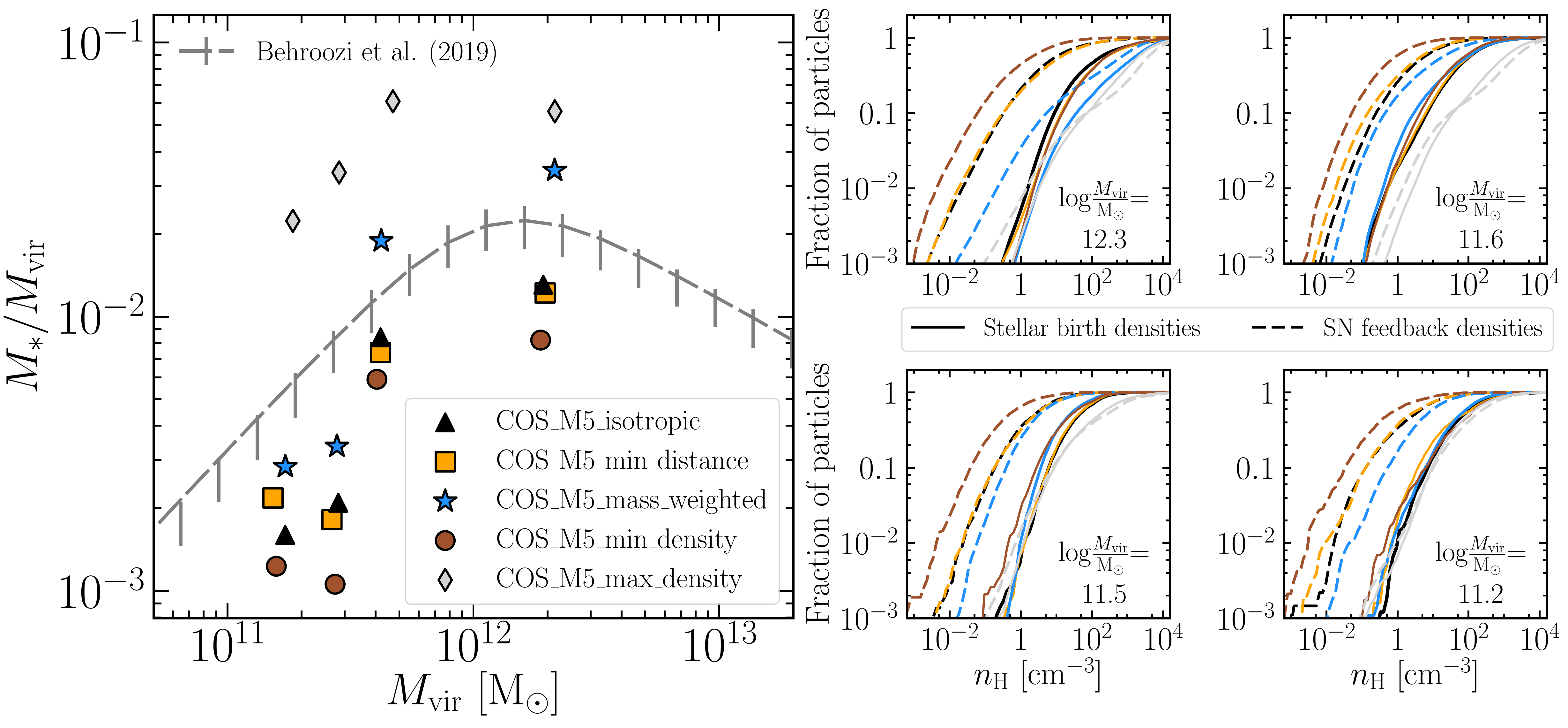} 
    \caption{\textit{Left:} Stellar-to-halo mass ratio versus halo mass shown for the four most massive central galaxies at $z=0$ in cosmological simulations of (6.25 cMpc)$^3$ volume at a gas mass resolution $m_{\rm gas} = 2.2 \times 10^{5} \, \rm M_\odot$, with five different ways of selecting gas particles that receive SN energy (colour-coded). The relation from the empirical model of \citet{2019MNRAS.488.3143B} is shown for reference only. \textit{Right:} Each smaller panel shows the cumulative distributions of stellar birth densities (\textit{solid}) and SN feedback densities (\textit{dashed}) for one of the four most massive galaxies from the left panel, for the same five methods of neighbour selection (colour-coding is the same as in the left panel). Consistently, the \textsf{min\_density} (\textsf{max\_density}) method yields the lowest (highest) stellar masses and the lowest (highest) feedback densities.}
    \label{fig:cosmo_main_smhm}
\end{figure*}

An ideal isotropic distribution would yield $\langle \cos \theta\rangle(z)=0.0$, while a negative slope with $z$ indicates a bias towards the disk. We see that among the three main runs, the \textsf{isotropic} model is closest to the ideal scenario, while the \textsf{mass-weighted} is farthest. The reason that the \textsf{isotropic} model also gives a slightly biased distribution is numeral sampling: if certain angular directions lack gas neighbours, there is no way that SN energy is injected there. In appendix \ref{appendix: isotropy} we show that the \textsf{isotropic} model approaches the ideal isotropic distribution even more closely with increasing resolution (and, hence, better sampling).

It should not be surprising that the distribution of the azimuthal angles in the \textsf{min\_distance} model is also not too far off from $\langle \cos \theta\rangle(z)=0.0$. By always selecting the closest gas neighbour for SN heating, we reduce the impact of density gradients within the SPH kernel, and hence the bias towards excessively injecting SN energy at higher densities. This reduction in the bias increases the likelihood that the closest neighbour can exist in any angular direction (relative to the star particle), which makes the model look more isotropic. This is also partly the reason why in the previous plots, the differences between \textsf{min\_distance} and \textsf{isotropic} were much smaller than between \textsf{mass-weighted} and \textsf{isotropic}. One of the statistics where \textsf{min\_distance} does (by construction) stand out from both \textsf{isotropic} and \textsf{mass-weighted} is the distribution of distances between star particles and their heated gas neighbours (see appendix \ref{appendix: distance_to_sn_events}).

\subsubsection{Resolution effects}
  
In Figure \ref{fig:star_formation_res}, we explore the impact of resolution on the variations in star formation rates caused by the differences in the adopted neighbour-selection algorithm. For clarity, we compare only the \textsf{isotropic} and \textsf{mass-weighted} models (the \textsf{min\_distance} model is again nearly indistinguishable from \textsf{isotropic}). We vary the resolution in the simulations by increasing and decreasing the gas-particle mass by a factor of $8$ relative to its fiducial value of $m_{\rm gas} = 10^{5} \, \rm M_\odot$. 

The difference in the star formation histories between the \textsf{isotropic} and \textsf{mass-weighted} methods decreases with decreasing resolution and nearly disappears for $m_{\rm gas} = 8 \times 10^{5} \, \rm M_\odot$. This is a direct consequence of the overall smoother distribution of gas at the lower resolution. The \textsf{isotropic} and \textsf{mass-weighted} methods both exhibit good convergence at our fiducial resolution $m_{\rm gas} = 10^{5} \, \rm M_\odot$ (relative to the higher, $m_{\rm gas} = 1.25 \times 10^{4} \, \rm M_\odot$ resolution), although the \textsf{isotropic} model converges slightly better at late times, $t\gtrsim 0.5$ Gyr.

\subsubsection{Impact of an effective pressure floor}
  
In Figure \ref{fig:star_formation_eos}, we show the impact of including an effective pressure floor $P_{\rm eos} \propto \rho_{\rm gas}^{4/3}$, which is normalised at density $n_{\rm H} = 0.1 \, \mathrm{cm}^{-3}$ to a temperature  $T = 8 \times 10^3 \, \mathrm{K}$. We plot galaxy star formation rates versus time, with (dashed) and without (solid) the effective pressure floor for the \textsf{isotropic} (black) and \textsf{mass-weighted} (blue) neighbour-selection methods (4 runs in total).

The differences in galaxy star formation rates between the \textsf{isotropic} and \textsf{mass-weighted} models are similar in the two cases, though they are slightly smaller if the pressure floor is included. Compared to runs without a pressure floor, the two runs with the pressure floor have smoother star formation histories and lack an initial burst of star formation. All these differences are due to the pressure floor yielding a smoother gas distribution. By the end of the simulations, the star formation rate in the \textsf{isotropic} model is a factor of 2 lower than for the \textsf{mass-weighted} model, nearly independent of the presence or absence of the pressure floor.

\subsection{Cosmological simulations}

We complete our analysis with the results from (6.25 cMpc)$^{3}$ cosmological volumes, at a gas mass resolution of  $m_{\rm gas} = 2.2 \times 10^{5} \, \rm M_\odot$. In essence, these confirm that our findings from the isolated galaxy runs presented above also apply to cosmological simulations. 

The left panel of Figure \ref{fig:cosmo_main_smhm} displays stellar-to-halo mass ratios for the four most massive central galaxies at $z=0$ versus halo mass, for the cosmological simulations with five neighbour-selection methods: \textsf{isotropic} (black triangles), \textsf{min\_distance} (orange squares), \textsf{mass-weighted} (blue stars), \textsf{min\_density} (brown circles), and \textsf{max\_density} (light-grey diamonds). For the halo masses we use the definition of the virial mass from \citet{1998ApJ...495...80B}; our stellar masses are computed in 3D apertures of radius $30$ kpc. For reference, we also show the median $z=0$ stellar-to-halo mass ratio taken from the empirical model of \citet{2019MNRAS.488.3143B}, with the error bars denoting the 16$^{\rm th}$ and 84$^{\rm th}$ percentiles.

Among the five runs, all galaxies -- regardless of their halo mass --  have the smallest (largest) stellar mass in the \textsf{min\_density} (\textsf{max\_density}) model, while the \textsf{min\_distance} and \textsf{isotropic} methods yield very similar stellar masses. These results are consistent with those from our isolated galaxy runs. Furthermore, the stellar masses of all four galaxies in the \textsf{mass-weighted} model are higher than those in the \textsf{isotropic} and \textsf{min\_distance} models by a factor of a few level, which is thus certainly not negligible. A similar shift in stellar masses is achieved by varying the SN energy budget by a factor of a few  \citep[see, e.g.][]{Crain2015}. 

In the four right, smaller panels in the right-hand half of Figure \ref{fig:cosmo_main_smhm}, we show  the cumulative distributions of stellar birth gas densities (solid) and SN feedback densities (dashed), in the four most massive galaxies from the left panel. The colour coding is the same as in the left panel. We see that the run with the \textsf{min\_density} (\textsf{max\_density}) algorithm, shown in brown (light-grey), has the lowest (highest) SN feedback densities in all four galaxies. As a result, SN feedback is most (least) effective in the simulation with the \textsf{min\_density} (\textsf{max\_density}) method and the galaxies in this model have the lowest (highest) star formation rates and, consequently, the lowest (highest) $z=0$ stellar masses. The feedback densities in the \textsf{mass-weighted} model are higher than those in  \textsf{min\_distance} and \textsf{isotropic} in all four galaxies. All these results are in line with our findings for the isolated galaxy runs (see Figure \ref{fig:density_main}).

\section{discussion}
\label{sec: discussion}

\subsection{Comparison with previous work}

As discussed in $\S$\ref{sec: introduction}, drastically different implementations of SN feedback are used by different research groups, including the so-called thermal, kinetic, decoupled, mechanical, stochastic, delayed-cooling and multi-phase subgrid models. Possibly because of the large differences in the internal design of the model, the effects of gas-neighbour selection in SN feedback have largely been overlooked so far. Only recently has this issue begun to receive some attention \citep{hopkinsfeedback2018,Smith2018,2019MNRAS.483.3363H}.

\citet{hopkinsfeedback2018} designed an algorithm to ensure statistical isotropy for their mechanical SN feedback model. In their algorithm, SN energy and momentum are distributed among gas elements neighbouring the stellar particle in proportion to the solid angles subtended by these elements, as seen by the star. The solid angles are estimated based on the effective surface areas shared by the star with its neighbours. \citet{hopkinsfeedback2018} ran zoom-in simulations of a Milky Way-like galaxy at gas-mass resolution $m_{\rm gas} = 7 \times 10^{3} \, \rm M_{\odot}$ using the mesh-free, Lagrangian code \textsc{gizmo} \citep{2015MNRAS.450...53H} in its finite-mass mode; they compared the isotropic algorithm to a naive, grid-aligned one, typical for Cartesian grid-based codes. They found that in the latter case, the galaxy has a different morphology, with a disk that is significantly more compact, and attributed this effect to the removal of angular momentum from recycling material due to the (forced) grid alignment.

\citet{Smith2018} implemented a mechanical subgrid model for SN feedback alongside the \citet{hopkinsfeedback2018} isotropic algorithm in the moving-mesh, quasi-Lagrangian code \textsc{arepo} \citep{2010MNRAS.401..791S} and ran simulations of an isolated galaxy with a total mass of $10^{10} \, \rm M_\odot$. They showed that at a resolution of $m_{\rm gas} = 2 \times 10^{3} \, \rm M_{\odot}$, choice of the SPH kernel weighting instead of the isotropic algorithm can result in unphysical shells propagating through the galaxy disk, sweeping up most of its gas mass. This is because if the SPH kernel weighting is used, most of the SN momentum is injected into the disk -- the region where most of the gas neighbours are -- rather than perpendicular to the disk.

\cite{2019MNRAS.483.3363H} ran simulations of dwarf galaxies at much higher resolution ($m_{\rm gas} = \rm 1 \, M_\odot$) and used a mechanical feedback model together with the \textsc{healpix} tessellation library \citep{2005ApJ...622..759G}, which makes the injection of SN energy and momentum isotropic. In \cite{2019MNRAS.483.3363H},
the $4\uppi$ solid angle was split into $12$ \textsc{healpix}  pixels; within each pixel, they looked for $8$ gas particles in which SN energy and momentum would be injected, therefore guaranteeing that the distribution of SN energy and momentum is isotropic at the pixel level for all possible distributions of gas neighbours. They found that the type of the injection scheme -- isotropic (with \textsc{healpix}) or naive (without \textsc{healpix}) -- makes negligible difference for their dwarf galaxies. Such an outcome is not necessarily in conflict with our findings (and the results from the other works) because the \citet{2019MNRAS.483.3363H} simulations had much higher resolution allowing the adiabatic phase of SN blastwaves to be fully resolved. Additionally, compared to the Milky Way-mass galaxy used in our work, in dwarf galaxies such as one used in \cite{2019MNRAS.483.3363H}, gradients in the gas density field are generally lower and the distribution of gas particles is generally smoother and more isotropic. Under such `less extreme' conditions, we expect the effects of neighbour selection to be less significant than those found in our work for the Milky Way-mass galaxy. Indeed, we have run simulations of a dwarf galaxy (not shown here) with $M_{\rm 200} = 1.37 \times 10^{10} \,\rm  M_\odot$,  $c=14.0$, and $m_{\rm gas} = 1.6\times 10^3 \, \rm M_\odot$ for the five models of neighbour selection and found that the effects of neighbour selection are strongly suppressed.

We note that our isotropic algorithm is fundamentally different from those in \citet{hopkinsfeedback2018} and \cite{2019MNRAS.483.3363H} because our method is based on rays, not on surface areas, \textsc{healpix} pixels, or solid angles; and because it is used together with the stochastic subgrid model where gas neighbours are heated \textit{individually}, not with the mechanical model where \textit{all} gas neighbours partake in feedback events. We emphasize that the \citet{hopkinsfeedback2018} and \cite{2019MNRAS.483.3363H} isotropic algorithms are superior to ours when used together with SN subgrid models like the mechanical model, in which the feedback energy is distributed among many ($N \gg 1$) gas neighbours and in all angular directions. This is mainly because our isotropic method is designed for a different class of SN models -- the models such as the \citet{2008MNRAS.383.1210S} kinetic model and the \citetalias{2012MNRAS.426..140D} thermal model where large amounts of energy are injected into one or a handful of gas neighbours. When applied to SN models distributing feedback energy among many neighbours, our method will be subject to substantial noise due to the stochastic sampling of angular directions by the finite number of rays. 

Another potential weakness of our isotropic method is that it may fail when certain angular directions lack gas neighbours, which is not the case in \cite{2019MNRAS.483.3363H}. This potential problem can be seen in Figure \ref{fig:isotropy_main} where our isotropic method is unable to produce a \textit{fully} isotropic distribution of the feedback energies, $\langle \cos \theta \rangle(z) = 0$, though it closely approaches it. The chance of lacking gas neighbours can be reduced by increasing the effective number of gas neighbours within the kernel or by adopting the neighbour finding strategy from \citet{hopkinsfeedback2018} where a star particle is allowed to inject SN energy not only into the neighbours within its own kernel but also into those gas elements that are outside the stellar kernel but whose (larger) kernels contain the star particle. Both improvements will come, however, with a certain computational cost and at low resolution may cause the SN energy to be distributed over unphysically large distances.

\subsection{Extension to other hydro solvers than SPH}

So far all discussion and results in this work have been presented in the context of SPH. In principle, our isotropic, ray-based algorithm from $\S$\ref{subsectuion isotropic model} can also be implemented in meshless codes like \textsc{gizmo} or moving-mesh codes like \textsc{arepo}. In the former, like in SPH solvers, the minimum arc length (equation \ref{eq:arclength}) should be computed between the rays and gas particles residing in the stellar kernel. In the latter, the arc length should be computed between the rays and the mesh-generation points of the neighbouring gas cells. We expect the qualitative results of this work to apply equally to these two set-ups. We also expect our \textsf{isotropic} model to look `more isotropic' than in this work if the average number of neighbouring gas elements is set to be greater than the value we use ($\approx 65$). That is because more neighbouring elements increases the probability of finding a particle close to the ray, resulting in a relatively better statistical isotropy of the model.

\subsection{Extension to other subgrid feedback models}

The neighbour-selection methods explored in this work can be used in combination with subgrid models for SN feedback other than the \citetalias{2012MNRAS.426..140D} thermal stochastic model. We expect the choice of the neighbour-selection algorithm to become less important if the subgrid model is less prone to radiative cooling losses. For example, in the subgrid models that inject SN energy in kinetic form and temporarily disable hydrodynamical forces acting on the kicked gas elements, the impact of the neighbour-selection strategy should be smaller than what we find here. This is because the kicked gas elements will freely escape from the ISM -- without suffering any energy losses -- regardless of the density of the gas they originate from. Conversely, in kinetic subgrid models for SN feedback where the kicked gas neighbours do lose energy to radiation and where the kick directions depend on the positions of the kicked gas particles with respect to the stars, the effect of neighbour selection may be even stronger than in our tests. This is because, depending on the adopted neighbour-selection method, not only the average densities of the kicked gas particles will differ, but also the paths that these particles will follow. 

We emphasize that in the tests of \cite{Smith2018}, it is the difference in the kick directions that dominates the changes in galaxy properties due to the injection scheme. In contrast, the effects of neighbour selection in our work are dominated by the differences in the densities at which SN energy is deposited. Because our SN feedback is thermal and stochastic, once a large amount of energy has been injected into a given gas neighbour, it will exert an isotropic pressure and the subsequent evolution of this energy will be governed by the hydro solver, which holds regardless of the neighbour selection method used in the simulation. We note also that for hydrodynamics schemes that do not smooth the gas density locally, changes in the method of neighbour selection may be more impactful than found for the density-energy SPH used in this work.

Finally, all neighbour-selection methods from $\S$\ref{sec: neighbour_selection} can be applied not only to subgrid models for SN type-II feedback but also to subgrid models for SN type-Ia feedback; subgrid models for stellar early feedback including HII regions, radiation pressure and AGB winds; and subgrid models for active galactic nucleus feedback from supermassive black holes. 

\section{Conclusions}
\label{sec: conclusions}

Although the role of supernova feedback in simulations of galaxy formation has been studied for many years, little attention has been given to the selection of the gas elements into which the SN energy is injected. In this work, we compared five methods of gas-element selection for SN feedback using isolated disk galaxy simulations as well as cosmological simulations. We focused on investigating the impact of the choice of the gas neighbours into which the SN energy is injected and kept all other aspects of the feedback model constant. We considered the (conventional) \textsf{mass-weighted} neighbour-selection method, a new \textsf{isotropic} method, and approaches where the SN energy is injected either into the closest, most dense, or least dense gas element. We kept the SN subgrid model itself fixed and used the stochastic thermal prescription from \citetalias{2012MNRAS.426..140D}, who originally used mass weighting to select the gas particles to heat. Our main findings are as follows:
    
\begin{itemize}

    \item In the presence of density gradients, the \textsf{mass-weighted} neighbour-selection algorithm used in \citetalias{2012MNRAS.426..140D} is biased towards heating gas at higher densities, which is due to the mass-weighted nature of SPH solvers. This bias is suppressed with our new algorithm, where the SN energy is distributed among gas elements as isotropically as the numerical resolution allows (Fig. \ref{fig:isotropy_main}).
    
    \item Using isolated galaxy simulations, we showed that changes in the way that gas neighbours are selected for SN feedback leads to significant variations in the galaxy star formation rates (Fig. \ref{fig:star_formation_main}), stellar birth densities (Fig. \ref{fig:density_main}), and wind mass loading factors (Fig. \ref{fig:mass_loadting_main}). These variations are mostly driven by the efficiency of SN feedback: this depends on the densities at which the gas is heated by SNe, which in turn is sensitive to the neighbour-selection method (Fig. \ref{fig:stellar_birth_density_stellar_feedback_density}).
    
    \item The most (least) efficient SN feedback model is the one using the \textsf{min\_density} (\textsf{max\_density}) algorithm, where SNe heat the least (most) dense gas, which suffers the least (most) from radiative cooling losses. The three main models -- \textsf{isotropic}, \textsf{min\_distance}, and \textsf{mass-weighted} -- are all bracketed by these two extreme cases. Among the main models, \textsf{mass-weighted} feedback is consistently less efficient than  \textsf{isotropic} feedback, while \textsf{min\_distance} feedback is similarly efficient (e.g. Fig. \ref{fig:star_formation_main}, \ref{fig:mass_loadting_main}).

    \item The differences in the galaxy properties caused by the usage of different neighbour-selection algorithms increase with increasing resolution. The \textsf{mass-weighted} and \textsf{isotropic} models both show good convergence at our fiducial resolution, with the \textsf{isotropic} method leading to a slightly better convergence at late times (Fig. \ref{fig:star_formation_res}).
    
    \item The above results are not sensitive to whether or not a pressure floor is used in the simulation. We confirmed that our results remain valid if we impose an equation of state, $P_{\rm eos}\propto \rho_{\rm gas}^{4/3}$, normalised to temperature $T= 8 \times 10^3 \, \mathrm{K}$ at density $n_{\rm H} = 0.1 \, \mathrm{cm}^{-3}$ (Fig. \ref{fig:star_formation_eos}).
    
    \item The results from the isolated galaxy simulations remain valid in cosmological simulations. In the cosmological simulations, the highest (lowest) $z=0$ galaxy stellar masses are reached with the \textsf{max\_density} (\textsf{min\_density}) methods, while the stellar masses in the \textsf{isotropic} and \textsf{min\_distance} models are very similar and are consistently lower than those in \textsf{mass-weighted} by up to a factor of a few (Fig. \ref{fig:cosmo_main_smhm}).
    
    \item Our results are broadly in line with the findings of \citet{hopkinsfeedback2018} and \citet{Smith2018} for higher-resolution simulations.
    
\end{itemize}

We conclude that the effects of isotropy in the SN feedback (and the choice of the element selection algorithm in general) are very important for simulations of galaxy formation and should be given at least equal consideration as other elements of the SN feedback model.

In closing, we note that if the locality of the SN feedback is important, the \textsf{min\_distance} method can be the preferred option, especially in light of its nearly equally good isotropic character as the (explicitly) `isotropic' method. However, in the test cases presented here the number of neighbours that receive SN energy is small. From additional experiments (not presented here) we found that, as expected, the differences between \textsf{isotropic} and \textsf{min\_distance} increase when the number of energy-injection events per star particle per time-step increases, with \textsf{isotropic} yielding more efficient feedback than \textsf{min\_distance}. Finally, we do not advocate using the \textsf{min\_density} and \textsf{max\_density} methods as these are extreme choices without a good physical motivation.

\section*{Acknowledgements}

This work used the DiRAC@Durham facility managed by the Institute for Computational Cosmology on behalf of the STFC DiRAC HPC Facility (www.dirac.ac.uk). The equipment was funded by BEIS capital funding via STFC capital grants ST/K00042X/1, ST/P002293/1, ST/R002371/1 and ST/S002502/1, Durham University and STFC operations grant ST/R000832/1. DiRAC is part of the National e-Infrastructure. EC was supported by the funding from the European Union's Horizon 2020 research and innovation programme under the Marie Sklodowska-Curie grant agreement No 860744 (BiD4BESt).YMB gratefully acknowledges funding from the Netherlands Organization for Scientific Research (NWO) through Veni grant number 639.041.751. This work is partly funded by Vici grant 639.043.409 from the Netherlands Organisation for Scientific Research (NWO). The research in this paper made use of the \textsc{swift} open-source simulation code (\url{http://www.swiftsim.com}, \citealt{2018ascl.soft05020S}) version 0.9.0. All galaxy images in this work (Figures \ref{fig:isolated_galaxy_morphology_gas_surface_density}, \ref{fig:mass_fraction_of_heated_gas}) were created using \textsc{py-sphviewer}  \citep{alejandro_benitez_llambay_2015_21703} and the data analysis was carried out with the help of \textsc{swiftsimio} \citep{Borrow2020simio}, \textsc{numpy} \citep{2020Natur.585..357H}, and \textsc{matplotlib} \citep{2007CSE.....9...90H}.

\section*{Data Availability}

The \textsc{swift} simulation code, including the implementation of the neighbour-selection methods for SN feedback explored in this work, is publicly available at \href{http://www.swiftsim.com}{http://www.swiftsim.com}. The data underlying this article will be shared on reasonable request to the corresponding author.



\bibliographystyle{mnras}
\bibliography{main} 




\appendix

\section{Degree of isotropy as a function of resolution}
\label{appendix: isotropy}

\begin{figure}
 \centering
 \includegraphics[width=0.49\textwidth]{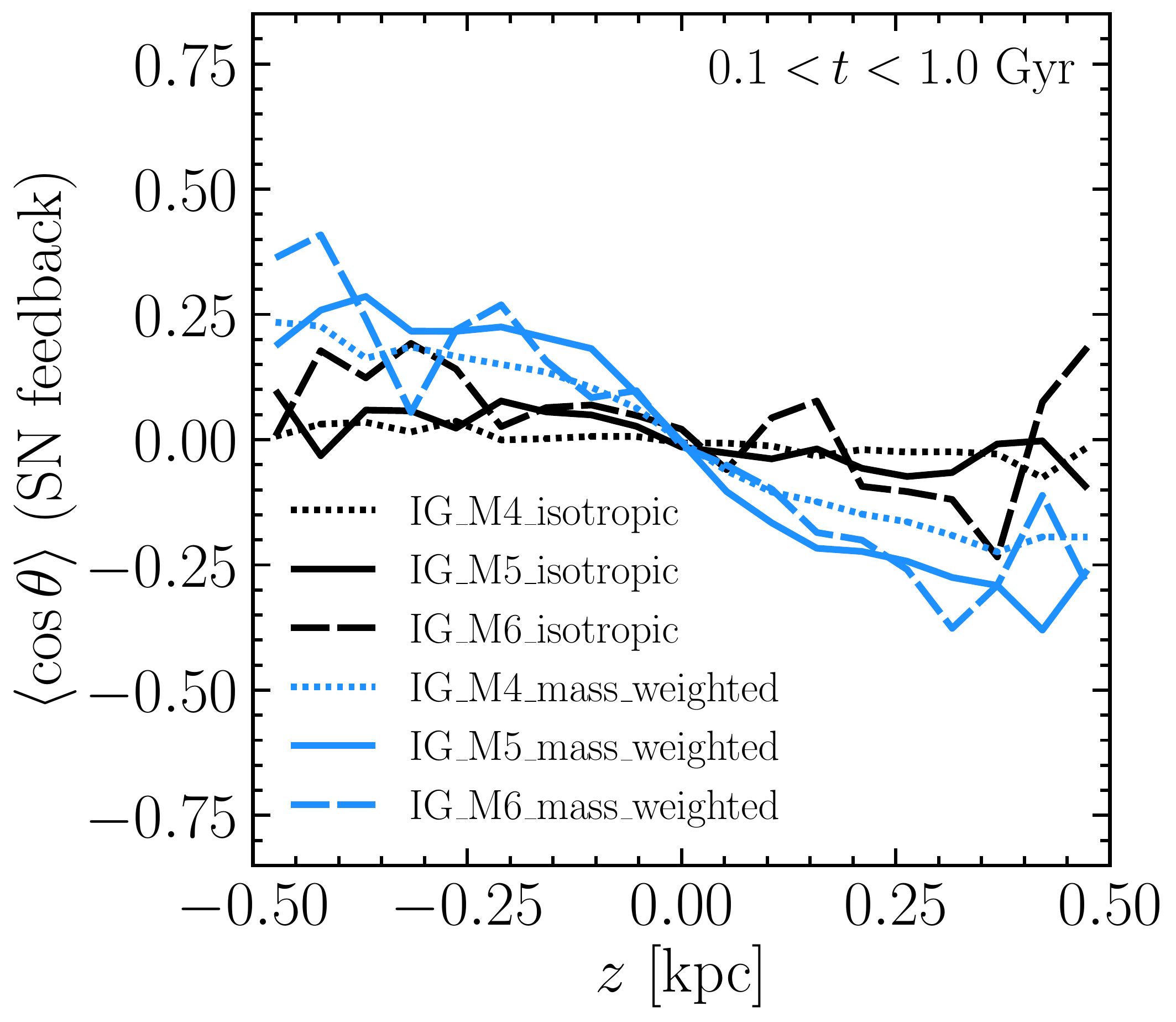}
 \caption{As Fig. \ref{fig:isotropy_main} but for the isolated galaxy simulations with the \textsf{isotropic} (\textit{black}) and \textsf{mass-weighted} (\textit{blue}) neighbour selection at three different resolutions: $m_{\rm gas} = 1.25 \times 10^{4} \, \rm M_\odot$ (\textit{short-dashed}), $m_{\rm gas} = 10^{5} \, \rm M_\odot$ (\textit{solid}), and $m_{\rm gas} = 8 \times 10^{5} \, \rm M_\odot$ (\textit{long-dashed}). The \textsf{isotropic} model approaches the ideal isotropic distribution $\langle \cos \theta \rangle(z) = 0$ with increasing resolution because of the better sampling of the gas density field.}
 \label{fig:iso_main_res}
 \end{figure}

In Figure \ref{fig:iso_main_res} we show the average cosine of the azimuthal angle in SN feedback defined in equation (\ref{eq:aver_cosine}), plotted against the height above the disk at which SN feedback events occur. The data is taken from the isolated galaxy simulations at three different resolutions,  $m_{\rm gas} = 1.25 \times 10^{4} \, \rm M_\odot$ (short-dashed), $m_{\rm gas} = 10^{5} \, \rm M_\odot$ (solid), and $m_{\rm gas} = 8 \times 10^{5} \, \rm M_\odot$ (long-dashed) for two neighbour-selection methods, \textsf{isotropic} (black) and \textsf{mass-weighted} (blue). 
 
As expected, the \textsf{isotropic} model yields a more isotropic distribution than the \textsf{mass-weighted} model, at all three resolutions. Moreover, the higher the resolution, the closer the \textsf{isotropic} model is to the ideal isotropic distribution $\langle \cos \theta \rangle(z) = 0$. This trend is caused by the better sampling of the gas density field with SPH particles when the resolution increases. 

\section{Distribution of distances between stars and their heated gas neighbours}
\label{appendix: distance_to_sn_events}

\begin{figure}
 \centering
 \includegraphics[width=0.45\textwidth]{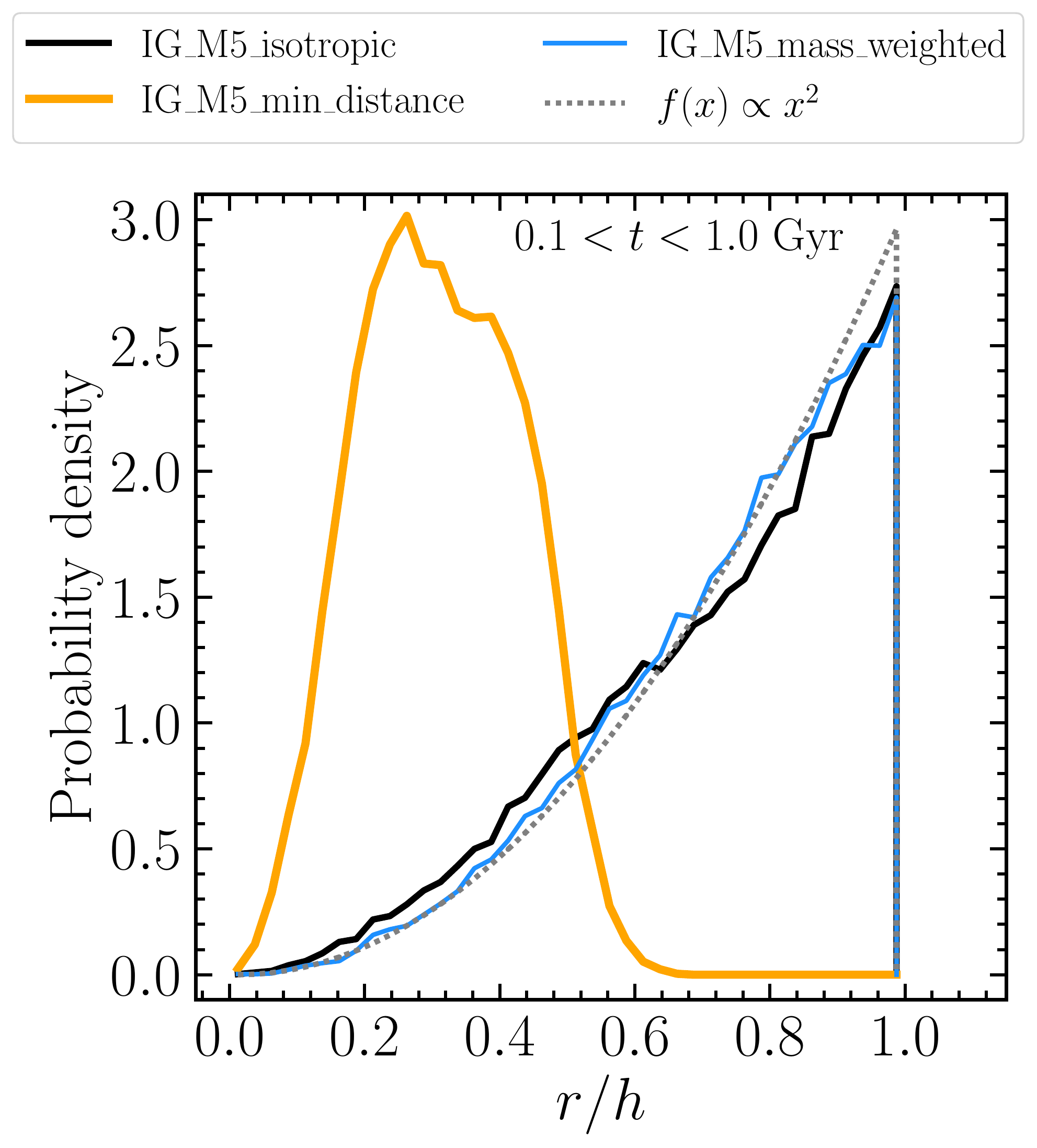}
 \caption{Probability density of the distances between stellar particles and the gas neighbours they heated in their last SN event, in units of the stellar particles' SPH smoothing length at the moment of the feedback, in the isolated galaxy simulations with resolution $m_{\rm gas} = 10^{5} \, \rm M_\odot$ with \textsf{isotropic} (\textit{black}), \textsf{min\_distance} (\textit{orange}) and \textsf{mass-weighted} (\textit{blue}) neighbour-selection methods. The SN events are selected from the time interval $0.1 < t < 1.0$ Gyr. The dashed grey curve gives the probability density for $f(x) \propto x^2$, corresponding to a radially unbiased distribution.}
 \label{fig:r_over_h_main}
 \end{figure}

In Figure \ref{fig:r_over_h_main} we show the probability density distribution of the distances between stellar particles and the gas neighbours they heated in their last supernova event, in the isolated galaxy simulations with resolution $m_{\rm gas} = 10^{5} \, \rm M_\odot$ with \textsf{isotropic} (black), \textsf{min\_distance} (orange) and \textsf{mass-weighted} (blue) neighbour-selection methods. The distances are normalised by the stellar particles SPH smoothing length at the moment of SN feedback. The SN events are selected from the time interval $0.1 < t < 1.0$ Gyr. For reference, we also show the probability density distribution of the form $f(x) \propto x^2$ (grey dashed). 

We find that the probability density distributions for \textsf{isotropic} and \textsf{mass-weighted} are both very close to the shape $f(x) \propto x^2$. This is expected because the SPH kernel is of spherical shape so that the (average) number of gas neighbours increases as $r^2$ where $r$ is the distance from the kernel's centre. The difference between \textsf{mass-weighted} and \textsf{isotropic} is only in how gas neighbours are selected based on their spherical angular coordinates. In contrast (but also as expected), the \textsf{min\_distance} model heats gas particles at much smaller distances. The probability density distribution for \textsf{min\_distance} peaks at the distance between stars and gas equal to $\approx 25$ per cent of the stars' smoothing length, and goes to zero at distances exceeding $\approx 70$ per cent of the smoothing length.
 
\bsp	
\label{lastpage}
\end{document}